\documentclass{aa}  
\usepackage{aalongtable}
\usepackage{graphicx}
\usepackage{natbib}
\bibpunct{(}{)}{:}{a}{}{,}
\usepackage{txfonts}
\begin{document}
   \title{Precise radial velocities of giant stars. III.\thanks{Based on data obtained at UCO/Lick Observatory, USA.}}

   \subtitle{Spectroscopic stellar parameters}

   \author{S. Hekker \inst{1} 
    \and J. Mel\'{e}ndez \inst{2} }

   \offprints{S. Hekker, \\
                    email: saskia@strw.leidenuniv.nl}
   
   \institute{Leiden Observatory, Leiden University, P.O. Box 9513, 2300 RA
Leiden, The Netherlands
                  \and
              Research School of Astronomy \& Astrophysics, Mount Stromlo Observatory, Cotter Road, Weston Creek, ACT 2611, Australia     
}

   \date{Received <date>; accepted <date>}

 
  \abstract
{A radial velocity survey of about 380 G and K giant stars is ongoing at Lick observatory. For each star we have a high signal to noise ratio template spectrum, which we use to determine spectroscopic stellar parameters. }
{The aim of this paper is to present spectroscopic stellar parameters, i.e. effective temperature, surface gravity, metallicity and rotational velocity for our sample of G and K giant stars. }
{Effective temperatures, surface gravities and metallicities are determined from the equivalent width of iron lines, by imposing excitation and ionisation equilibrium through stellar atmosphere models.  Rotational velocities are determined from the full width at half maximum (FWHM) of moderate spectral lines. A calibration between the FWHM and total broadening (rotational velocity and macro turbulence) is obtained from stars in common between our sample and the sample from Gray (1989). Macro turbulence is determined from the macro turbulence vs. spectral type relations from Gray (2005). }
{The metallicity we derive is essentially equal to the literature values, while the effective temperature and surface gravity are slightly higher by 56 K and 0.15 dex, respectively. A method comparison is performed with 72 giants in common with Luck and Heiter (2007), which shows that both methods give similar results. Our rotational velocities are comparable with those obtained by Gray (1989), but somewhat higher than those of de Medeiros \& Mayor (1999), which is consistent with the different diagnostics used to determine them.}
{We are able to determine spectroscopic stellar parameters for about 380 G and K giant stars uniformly (112 stars are being analysed spectroscopically for the first time). For stars available in the literature, we find reasonable agreement between literature values and values determined in the present work. In addition, we show that the metallicity enhancement of companion hosting stars might also be valid for giant stars\textbf{\rm{, with the planet hosting giants being 0.13 $\pm$ 0.03 dex (i.e.~ 35 $\pm$ 10\%) more metal rich than our total sample of stars.}}}

   \keywords{stars: abundances -- stars: fundamental parameters -- stars: rotation -- methods: observational -- techniques: spectroscopic
               }
 \authorrunning{ S. Hekker \& J. Mel\'{e}ndez}
  \maketitle
%
\section{Introduction} 
For the determination of spectroscopic stellar parameters, one needs high resolution spectra with high signal to noise ratio. These spectra are available from radial velocity surveys and are often used to determine stellar parameters. For instance, properties of cool stars from the Keck, Lick and AAT planet search are described by \citet{valenti2005}. Atmospheric parameters for stars observed by the N2K consortium \citep{fischer2005a} are described by \citet{robinson2007}. \citet{santos2004,santos2005} present stellar parameters and metallicities from the planet search using ESO facilities and the ELODIE spectrograph at the 1.93 m telescope at the Observatoire de Haute Provence. Also, basic stellar parameters for 72 evolved stars, previously studied for radial velocity variations, are presented by \citet{dasilva2006}. Some of these results are not only interesting in terms of the stellar parameters, but also reveal which stars are most likely to harbour sub-stellar companions. As first shown by \citet{gonzalez1997}, and confirmed with larger samples by \citet{fischervalenti2005} and \citet{santos2005}, metal rich stars are more likely to harbour companions than metal poor ones. 

Spectroscopic stellar parameters are most commonly determined by fitting the observed spectrum directly, see for instance \citet{valenti2005}, or by imposing excitation and ionisation equilibrium for metal lines, using an LTE analysis and a grid of model atmospheres, see for instance \citet{santos2004,santos2005}, \citet{dasilva2006}, \citet{takeda2002} and \citet{luck2007}.

Rotational velocity and macro turbulence can only be determined directly with the Fourier transform technique, see for instance \citet{gray1989}. \citet{benz1981} have shown that accurate rotational velocities can also be deduced for dwarfs from a cross correlation function, by performing a calibration with the direct measurements of \citet{gray1989}. \citet{demedeiros1999} extended this technique for giant stars. \citet{fekel1997} used the full width at half maximum (FWHM) of weak to moderate spectral lines to determine rotational velocities, also by performing a calibration with the results of \citet{gray1989}.

In 1999, a radial velocity survey of about 180 K giant stars was started at UCO/Lick Observatory, USA. This ongoing survey has recently been expanded to about 380 G and K giants.
From the initial sample of 180 stars, companions have been announced for $\iota$ Draconis \citep{frink2002} and Pollux \citep{reffert2006}. Stars with radial velocity variations of less than 20 ms$^{-1}$ were presented as stable stars by \citet{hekker2006a}, and an investigation into the mechanism(s) causing the radial velocity variations is presented by \citet{hekker2007}. Some binaries discovered with this survey, as well as an extensive overview of the sample, will be presented in forthcoming papers. 

In this paper, we determine stellar parameters, i.e. effective temperature (T$_{\rm{eff}}$), surface gravity ($\log$~g) and metallicity ([Fe/H]), as well as rotational velocity ($\varv \sin i$) for all stars in the sample. In Sect. 2, we describe the observations. In Sects. 3 and 4, we present the methods used, and results for the stellar parameters and rotational velocity, respectively. In Sect. 5 a summary of our results is presented.

\section{Observations}
For the radial velocity survey, giants were selected from the Hipparcos catalog \citep{esa1997}, based on the criteria described by \citet{frink2001}. The selected stars are all brighter than 6~mag, are presumably single and have photometric variations $< 0.06$~mag. These criteria are the same for the initial sample (K1 and later giants) as well as for the extension (G and K0, K1 giants). The survey started in 1999 at Lick observatory using the Coude Auxiliary Telescope (CAT) in conjunction with the Hamilton echelle spectrograph (R = 60\,000). The radial velocity measurements are performed with an iodine cell in the light path as described by \citet{marcy1992} and \citet{valenti1995}. Radial velocities are determined from the comparison of a stellar spectrum obtained with an iodine cell in the light path, and the convolution of a template iodine spectrum and a template stellar spectrum obtained without an iodine cell in the light path \citep{butler1996}.  For each target star we have a high signal to noise ratio template spectrum. These templates are used for the determination of the stellar parameters described in this paper.
Thorium-Argon images taken at the beginning and end of each night are used for wavelength calibration.

\section{Effective temperature, surface gravity, and metallicity}

\begin{table}
\begin{minipage}[t]{\linewidth}
\caption{Iron lines considered in our analysis.}
\label{Felines}
\centering
\begin{tabular}{lcrl}
\hline\hline
Ion & $\lambda$ [\AA] & $\chi$ [eV] & $\log$~gf\\
\hline
Fe I & 5775.080 & 4.220& $-$1.30\\
Fe I & 5848.129 & 4.607 & $-$0.9\\
Fe I & 5902.473 & 4.593 & $-$1.75\\
Fe I & 5916.247 & 2.453 & $-$2.99\\
Fe I & 6027.050 & 4.076 & $-$1.3\\
Fe I & 6093.644 & 4.607 & $-$1.41\\
Fe I & 6096.665 & 3.984 & $-$1.81\\
Fe I & 6098.244 & 4.558 & $-$1.8\\
Fe I & 6120.249 & 0.915 & $-$5.95\\
Fe I & 6151.618 & 2.176 & $-$3.30\\
Fe I & 6187.990 & 3.943 & $-$1.65\\
Fe I & 6240.646 & 2.223 & $-$3.39\\
Fe I & 6498.939 & 0.958 & $-$4.70\\
Fe I & 6574.228 & 0.990 & $-$5.00\\
Fe I & 6703.567 & 2.759 & $-$3.15\\
Fe I & 6725.357 & 4.103 & $-$2.30\\
Fe I & 6726.666 & 4.607 & $-$1.17\\
Fe I & 7421.558 & 4.638 & $-$1.80\\
Fe I & 7547.896 & 5.099 & $-$1.10\\
Fe I & 7723.208 & 2.279 & $-$3.62\\
Fe II & 5264.812 & 3.230 & $-$3.13\\
Fe II & 5425.257 & 3.200 & $-$3.22\\
Fe II & 6247.557 & 3.892 & $-$2.30\\
Fe II & 6369.462 & 2.891 & $-$4.11\\
Fe II & 6432.680 & 2.891 & $-$3.57\\
Fe II & 6456.383 & 3.904 & $-$2.05\\
\hline
\end{tabular}
\end{minipage}
\end{table}

Spectroscopic stellar parameters (T$_{\rm{eff}}$, $\log$~g and [Fe/H]) are determined by measuring the equivalent width (EW) of iron lines. The iron lines used in this work are listed in Table~\ref{Felines}. The lines were carefully selected to avoid blends by atomic and CN lines. CN blends were visually inspected by comparing a synthetic spectrum computed with laboratory CN lines \citep{melendez1999} with the high resolution visible atlas of the cool giant Arcturus \citep{hinkle2000}. The $\log$~gf values are based on laboratory works, in some cases with small adjustments using the Arcturus atlas. For Fe I, they are from the Oxford group \citep[e.g.][]{blackwell1995}), Hannover group \citep[e.g.][]{bard1994}), \citet{obrian1991}, \citet{may1974} and \citet{milford1994}. For Fe II, the $\log$~gf values are from the laboratory normalisation performed by \citet{melendez2006}.

\begin{figure}
\centering
\includegraphics[width=\linewidth]{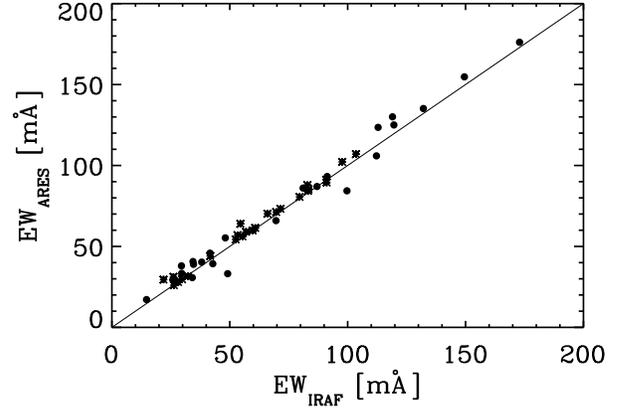}
\caption{EWs obtained with the ARES software as a function of EWs computed using IRAF. The dots indicate a star with T$_{\rm{eff}}$ = 4050 K (`cool') and the asterisks a star with T$_{\rm{eff}}$ = 4900 K (`hot'). The solid line is a 1 to 1 relation.}
\label{EWAI}
\end{figure}

It is very time consuming to determine EWs for about 380 stars by hand, using for instance the ``splot'' routine from IRAF\footnote{IRAF is distributed by National Optical Astronomy Observatories, operated by the Association of Universities for Research in Astronomy, Inc., under contract with the National Science Foundation, U.S.A.}, we therefore used the publicly available Automatic Routine for line Equivalent widths in stellar Spectra (ARES) \citep{sousa2007}. 
To check for possible differences between the EWs determined with ARES and those obtained with IRAF, we plot the EWs obtained with ARES vs. those obtained using IRAF (see Fig.~\ref{EWAI}). This comparison is done for a `hot'  (T$_{\rm{eff}}$ = 4900 K) and a `cool' (T$_{\rm{eff}}$ = 4050 K) star in the sample. The mean differences between the EWs measured with ARES and IRAF are: $\langle \rm EW_{ARES} - \rm EW_{IRAF} \rangle$ = 1.6 m\AA~and 2.2 m\AA, with standard deviations of 6.6 m\AA~and 2.6 m\AA, for the `cool' and `hot' star respectively. The nearly 1 to 1 relation between the EWs obtained with both methods shows that it is reasonable to use EWs obtained with ARES.
For some stars one or more lines appeared to be too strong (stronger than 200 m\AA) for a reliable parameter estimate. These lines are discarded. 

From the EWs, stellar parameters are determined by imposing excitation and ionisation equilibrium through stellar atmosphere models. The micro turbulence (v$_{t}$) was obtained by requiring no dependence of Fe I against equivalent width. We performed a spectroscopic LTE analysis using the 2002 version of MOOG \citep{sneden1973} and Kurucz model atmospheres, which include overshooting \citep{castelli1997}. The resulting stellar parameters for each star are listed in Table~\ref{parameters} (only available in the online version).  The reference solar iron abundance used in this study is A(Fe)$_{\sun}$ = 7.49 and was obtained using the same grid of Kurucz models. 

\begin{table}
\begin{minipage}[t]{\linewidth}
\caption{Internal errors due to changes in stellar parameters for three stars with different temperatures. The total error in the last column includes the observational errors (0.024 dex for Fe I and 0.037 dex for Fe II).}
\label{errors}
\centering
\begin{tabular}{lllll}
\hline\hline
Ion & $\Delta$ T$_{\rm eff}$ & $\Delta$ $\log$~g & $\Delta$ v$_{t}$ & $\sqrt{\Sigma x^{2}}$ \\
 & +80 K & $+$0.2 dex & $+$0.2 km\,s$^{-1}$ &\\
 \hline
HD156681 & (T = 4170 K) &&&\\
Fe I & -0.01 & +0.07 & -0.07  & 0.10\\
Fe II & -0.15 & +0.16 & -0.03  & 0.22\\
HD214868 &(T = 4445 K) &&&\\
Fe I & +0.01 & +0.07 & -0.06  & 0.10\\
Fe II & -0.12 & +0.18 & -0.04  & 0.22\\
HD165634& (T = 4980 K) &&&\\
Fe I & +0.06 & +0.02 & -0.06 & 0.09\\
Fe II & -0.06 & +0.10 & -0.07 & 0.14\\ 
\hline
\hline
\end{tabular}
\end{minipage}
\end{table}

Based on the scatter of the Fe lines, we obtained observational errors (standard errors) of 0.024 dex for Fe I and 0.037 dex for Fe II. Furthermore, we compute internal errors for a change of $+$80K, $+$0.2 dex and $+$0.2 km\,s$^{-1}$ in T$_{\rm{eff}}$, $\log$~g and micro turbulence, respectively. The errors due to change in stellar parameters are shown for three stars with different temperatures in Table~\ref{errors}.

\subsection{Comparison with the literature}

\begin{figure*}
\begin{minipage}{0.33\linewidth}
\centering
\includegraphics[width=\linewidth]{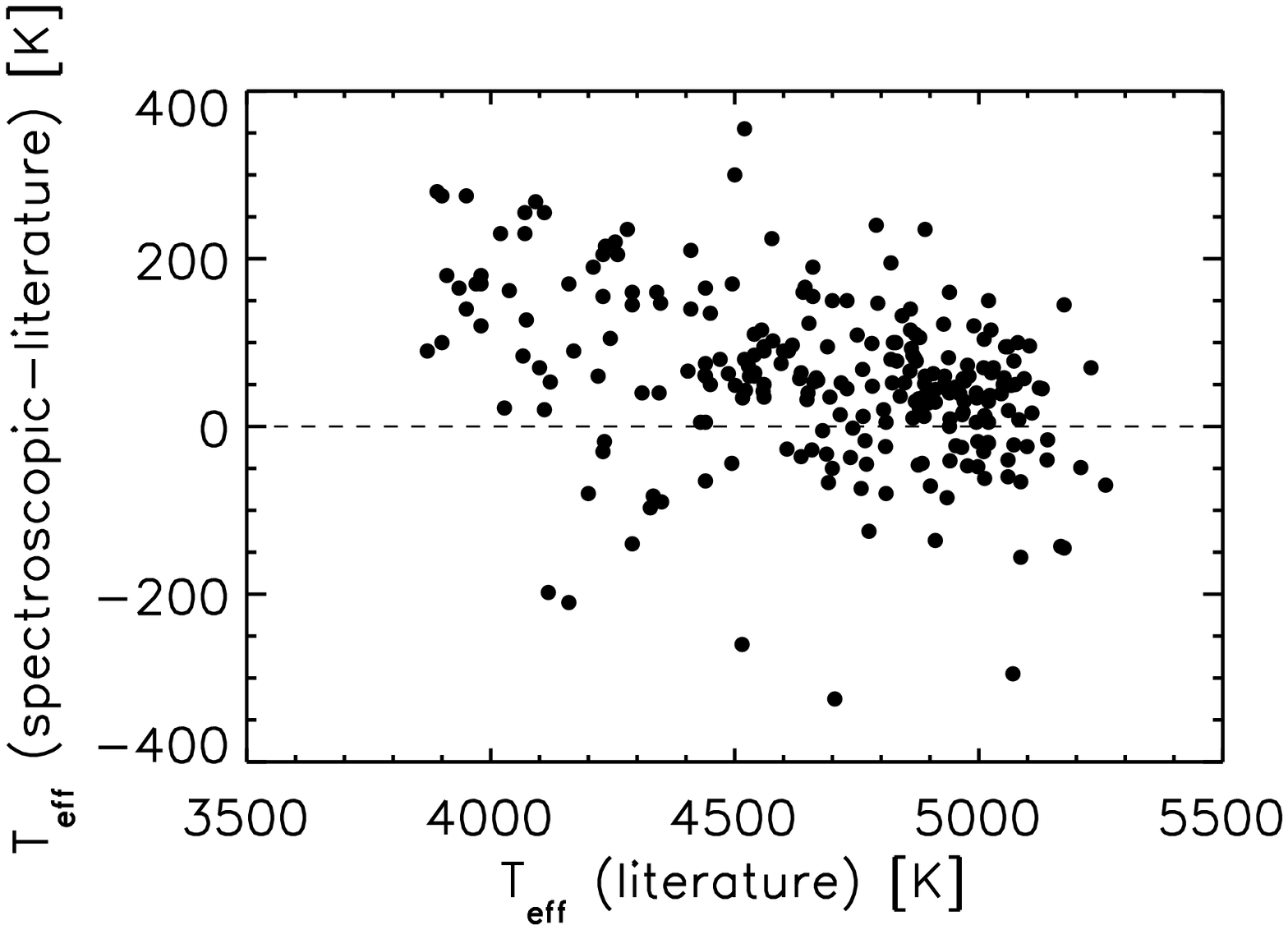}
\end{minipage}
\hfill
\begin{minipage}{0.33\linewidth}
\centering
\includegraphics[width=\linewidth]{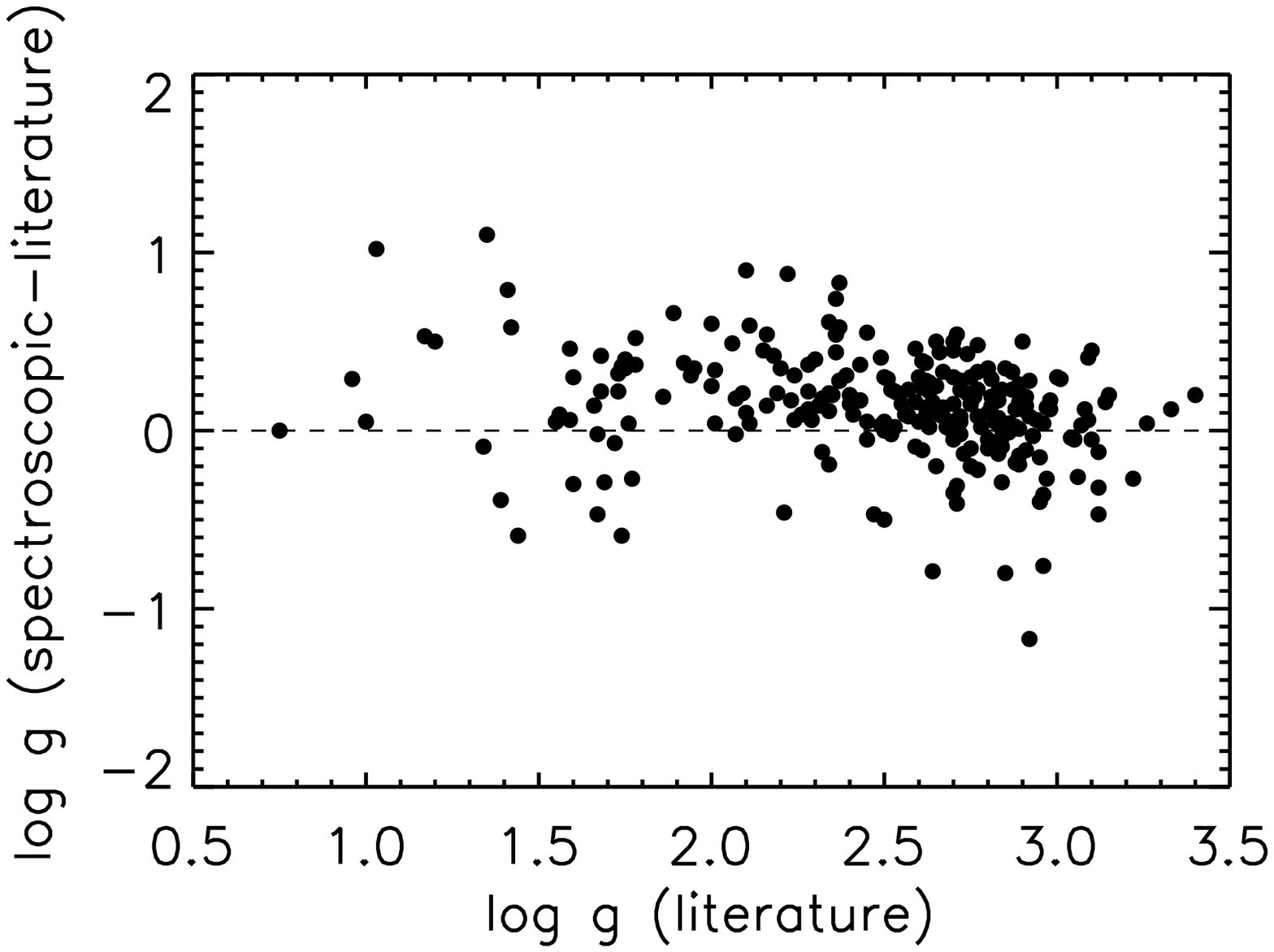}
\end{minipage}
\hfill
\begin{minipage}{0.33\linewidth}
\centering
\includegraphics[width=\linewidth]{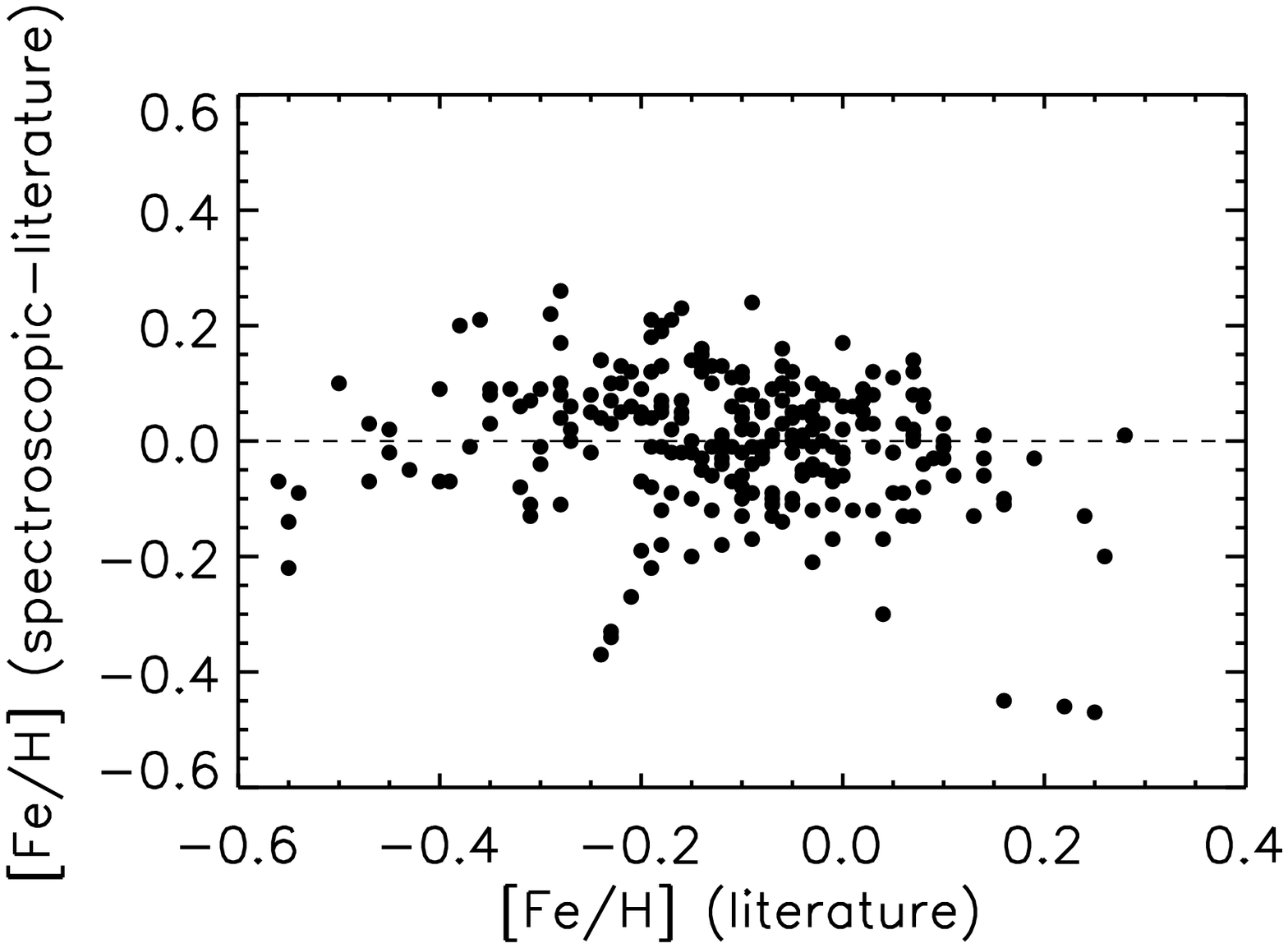}
\end{minipage}
\caption{Difference between our spectroscopic values and literature values as a function of literature values for the effective temperature (left), the logarithm of the surface gravity in cgs units (centre) and metallicity (right).}
\label{literature}
\end{figure*}

We compare our spectroscopic stellar parameters with values obtained from the literature. These literature values are from an updated version \citep{ramirez2005a} of the \citet{cayreldestrobel2001} [Fe/H] catalogue (254 stars in common), including the \citet{luck2007} catalogue. If possible, suspicious literature values were corrected according to the normalisation suggested by \citet{taylor1999}. In Fig.~\ref{literature} the difference between our spectroscopic and literature values are plotted for the temperature, surface gravity and metallicity. We find the following trimean difference and pseudo-sigma  for the stellar parameters:
\begin{eqnarray}
<[\rm{Fe/H}]^{spec} - [\rm{Fe/H}]^{lit}> = 0.01 \mathrm{ dex} & & \sigma=0.10 \mathrm{ dex} \nonumber\\
<\log g^{spec} - \log g^{lit}> = 0.15 \mathrm{ dex} & & \sigma=0.22 \mathrm{ dex} \nonumber\\
<T_{\rm{eff}}^{spec} - T_{\rm{eff}}^{lit}> = 56 \mathrm{ K} & & \sigma=84 \mathrm{ K} \nonumber
\end{eqnarray}
The trimean T is a robust estimate of central tendency: T=(Q1+2 x median+Q3)/4 where Q1 and Q3 are the first and third quartile. The pseudo-standard deviation $\sigma$ is obtained from the quartile deviation QD = (Q3-Q1)/2, employing $\sigma$ = 3/2 QD \citep{melendez2006}.

The difference in our spectroscopic metallicity and the literature value is essentially zero and we conclude that our metallicity scale is correct.
Furthermore, our T$_{\rm{eff}}$ values are in good agreement with the literature, with a scatter of only 84 K, and a zero point difference of 56 K, our T$_{\rm{eff}}$ values being higher than the values in the literature. From the left panel of Fig.~\ref{literature} one can see that the difference is largest for the coolest stars in the sample. This might be due to the fact that the models are less accurate for low temperatures. In addition, the number of spectral lines increases with decreasing temperature, the spectra might be too crowded at lower temperatures, and also the lines get stronger and more dependent on the micro turbulence. Our results below 4000 K should be interpreted with caution.

The spectroscopic gravities we derived also agree well with the literature, with a scatter of only 0.22 dex, and a zero point difference of 0.15 dex.
We checked whether the enhanced $\log$~g values from our spectroscopic analyses are related to the higher temperatures we obtained, compared to literature values. We therefore performed a test for three stars with T$_{\rm{eff}}$ 4170 K, 4445 K and 4980 K, respectively. We increased T$_{\rm{eff}}$ with 100 K  and determined $\log$~g, while keeping the micro turbulence fixed. For all three stars we obtained higher $\log$~g values for the increased temperatures. This reveals that the higher values for $\log$~g, compared to the literature values, are related to the higher effective temperatures.

\subsection{Comparison with \citet{luck2007}}

\begin{figure*}
\begin{minipage}{0.33\linewidth}
\centering
\includegraphics[width=\linewidth]{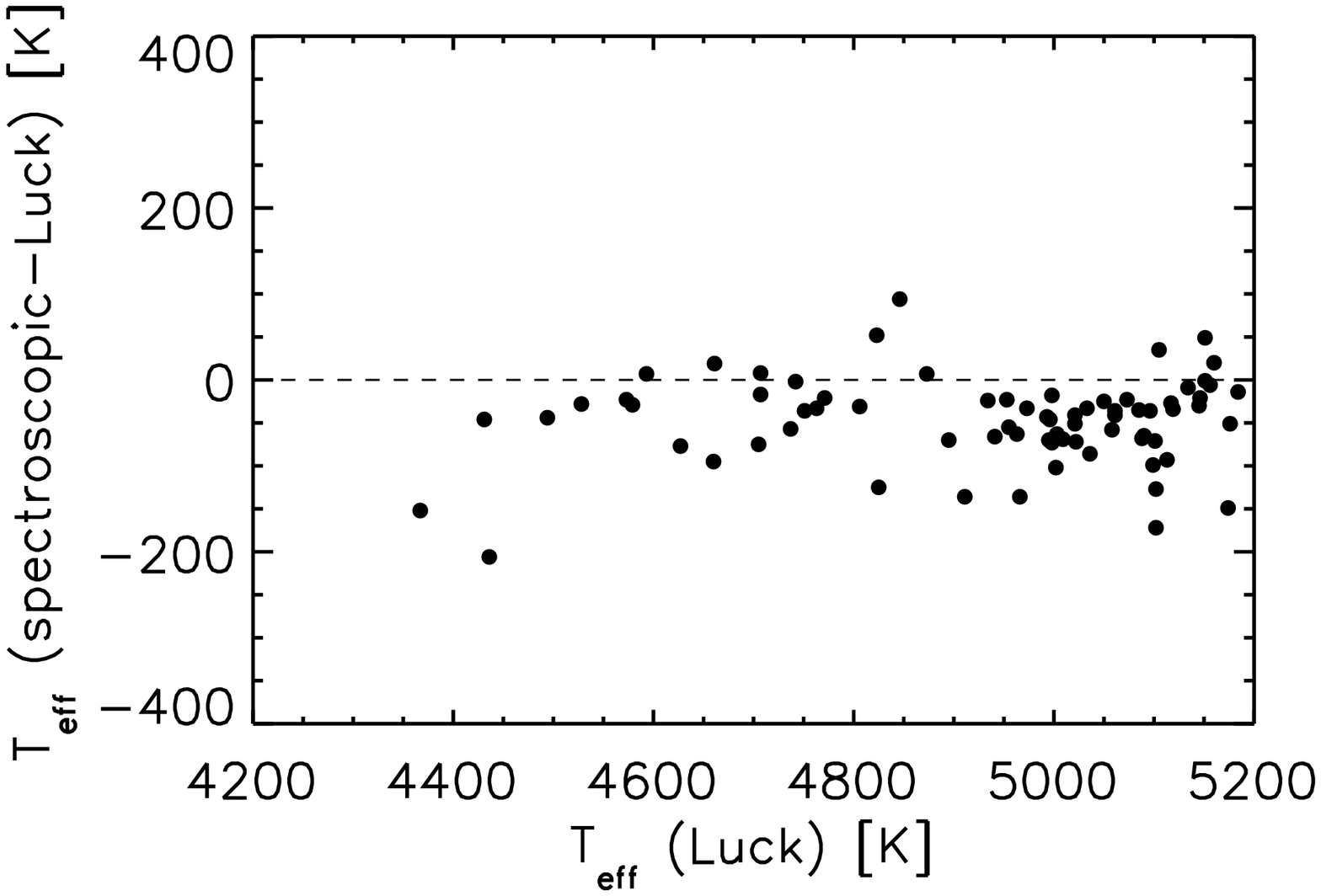}
\end{minipage}
\hfill
\begin{minipage}{0.33\linewidth}
\centering
\includegraphics[width=\linewidth]{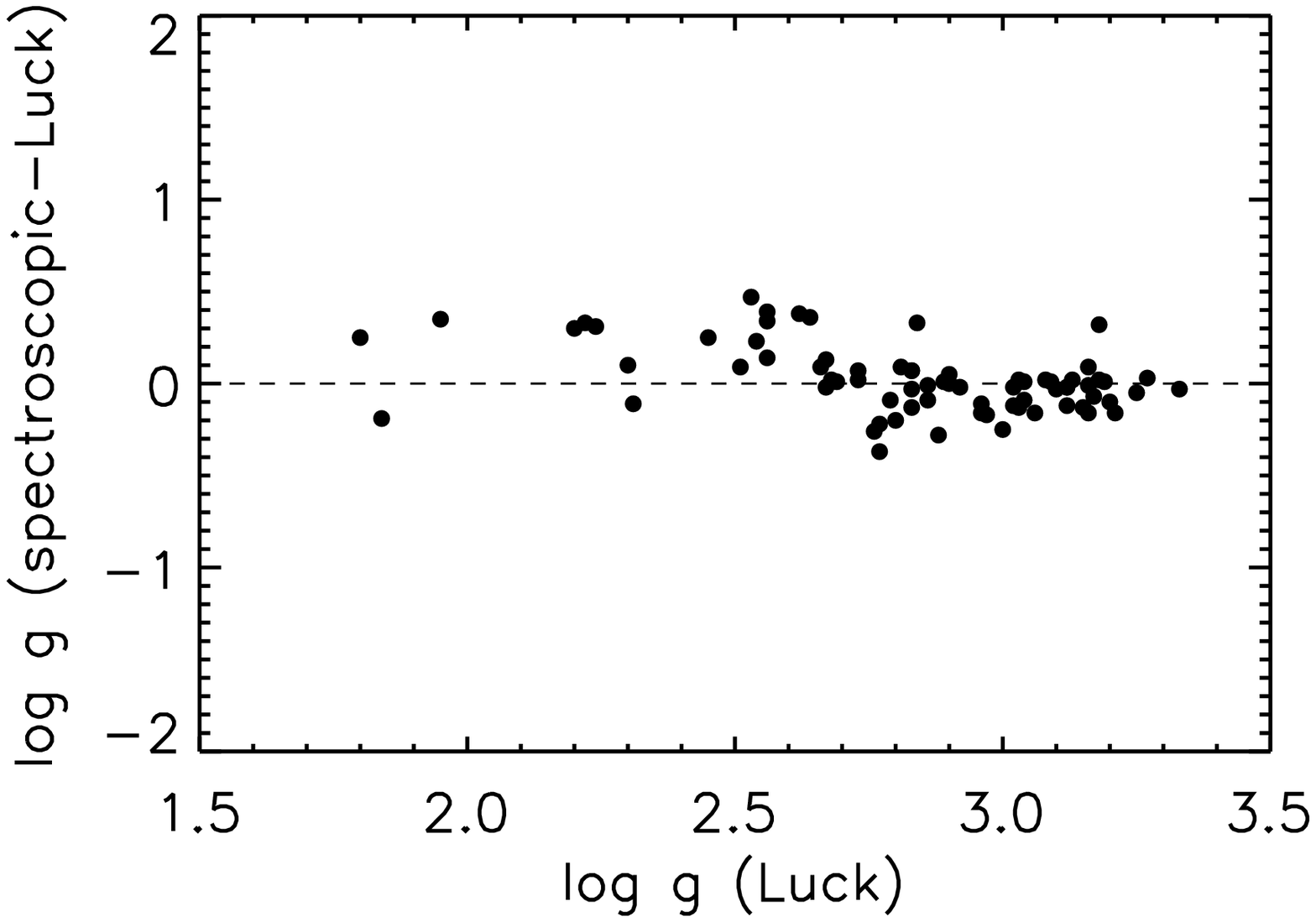}
\end{minipage}
\hfill
\begin{minipage}{0.33\linewidth}
\centering
\includegraphics[width=\linewidth]{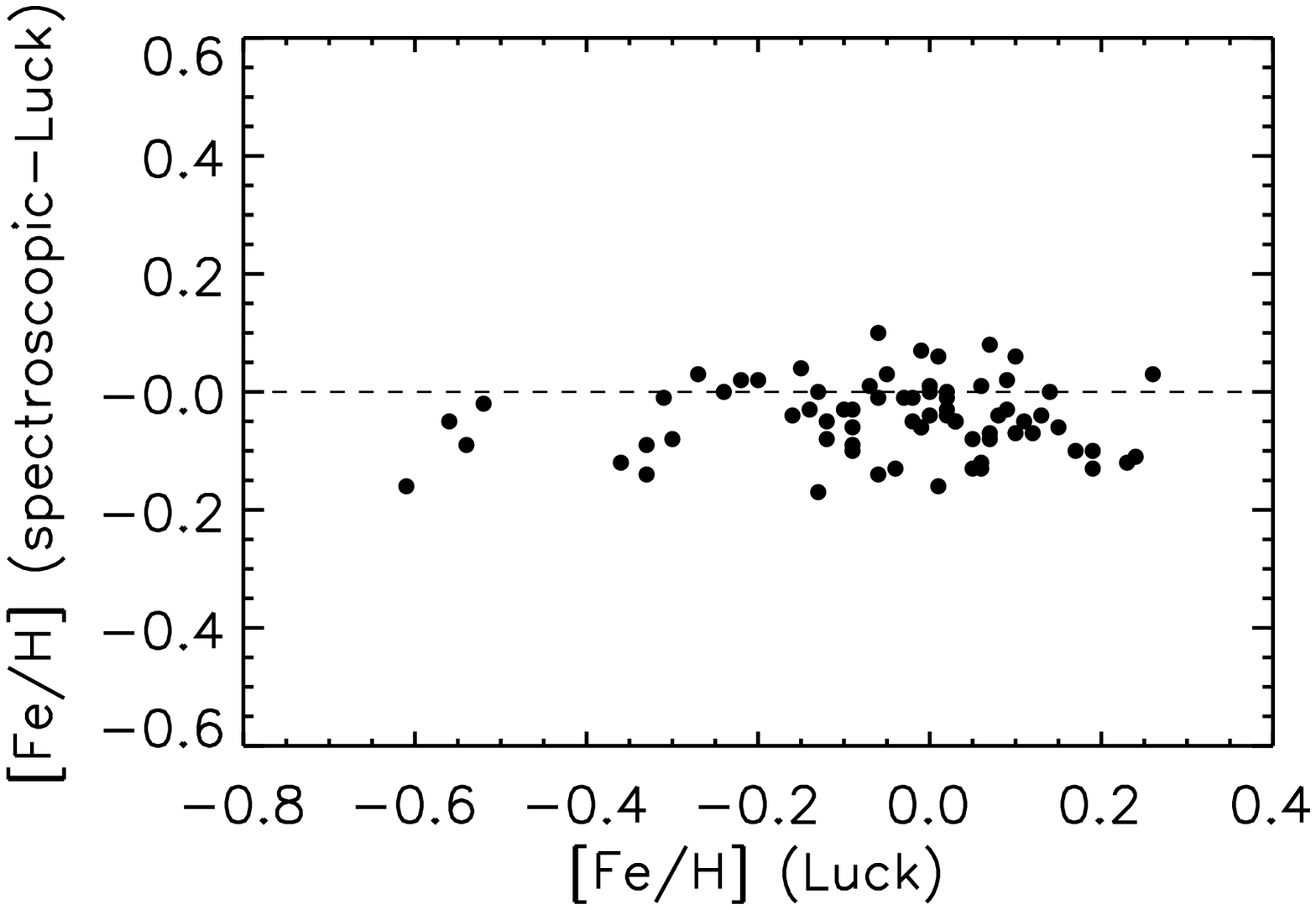}
\end{minipage}
\caption{Difference between our spectroscopic values and values from \citet{luck2007} as a function of \citet{luck2007} values for the effective temperature (left), the logarithm of the surface gravity in cgs units (centre) and metallicity (right) for the 72 stars in common between the samples. }
\label{luck}
\end{figure*}

Recently, \citet{luck2007} presented a homogeneous spectroscopic analysis of 298 giants in the local region, using between 300 and 400 Fe I lines for each star and MARCS stellar models \citep{gustafsson2003} on spectra with R=60\,000. We used the 72 stars in common between \citet{luck2007} and our sample, to see how well we can determine spectroscopic stellar parameters with just two dozen carefully selected iron lines instead of a few hundred iron lines. Note that different models are used for the two analyses, however spectra have the same resolution. A comparison for each parameter is shown in Fig.~\ref{luck}.
We find the following trimean difference and pseudo-sigma:
\begin{eqnarray}
<[\rm{Fe/H}]^{spec} - [\rm{Fe/H}]^{Luck}> = -0.05 \mathrm{ dex} & & \sigma=0.06 \mathrm{ dex} \nonumber\\
<\log g^{spec} - \log g^{Luck}> = 0.0 \mathrm{ dex} & & \sigma=0.15 \mathrm{ dex} \nonumber\\
<T_{\rm{eff}}^{spec} - T_{\rm{eff}}^{Luck}> = -43 \mathrm{ K} & & \sigma=35 \mathrm{ K} \nonumber
\end{eqnarray}
Our spectroscopic values are in good agreement with those obtained by \citet{luck2007}, with a scatter of only 0.06 dex for [Fe/H], 0.15 dex for $\log$~g and 35 K for the effective temperature. The mean difference in $\log$~g values is zero, while our metallicities and temperatures are slightly lower than those reported by \citet{luck2007}. These are probably systematic differences between both methods, because the pseudo-sigmas are relatively small. \citet{luck2007} have benchmarked their codes against Kurucz's WIDTH and SYNTHE codes and claim that all codes yield the same result to within expected numerical accuracy and differences due to different assumptions, primarily partition functions and damping. It is therefore likely that the different adopted model atmospheres (MARCS vs. Kurucz) and different $\log$~gf values cause the small systematic difference. \citet{luck2007} did not publish their line list and $\log$~gf values. Since they have used a much larger number of iron lines a comparison with our $\log$~gf values is probably not meaningful.

\subsection{Metallicity in companion hosting giants}

\begin{table*}
\begin{minipage}[t]{\linewidth}
\caption{Companion hosting giants with their mean metallicities and literature sources.}
\label{plgiants}
\centering
\begin{tabular}{llrl}
\hline\hline
HIP & HD & [Fe/H] & literature\\
\hline
4297 & 5319 & $+$0.04 & \citet{robinson2007b}\\
8928 & 11977 & $-$0.19 & \citet{dasilva2006}, \citet{sousa2006}\\
10085 & 13189 & $-$0.57 & \citet{schuler2005}, \citet{sousa2006}\\
 & 17092 & $+$0.19 & \citet{niedzielski2007}\\
19921 & 27442 & $+$0.33 & \citet{santos2003}, \citet{valenti2005}\\
20889 & 28305 & $+$0.10 & \citet{mishenina2006}, \citet{schuler2006}, \citet{sato2007}, this work\\
31688 & 47536 & $-$0.64 & \citet{sadakane2005}, \citet{dasilva2006}\\
36616 & 59686 & $+$0.11 & \citet{santos2005}, \citet{sadakane2005}, \citet{mishenina2006}, this work\\
37826 & 62509 & $+$0.06 & \citet{sadakane2005}, \citet{luck2007}, this work\\
42527 & 73108 & $-$0.24 & \citet{luck2007}, \citet{dollinger2007}\\
58952 & 104985 & $-$0.29 & \citet{santos2005}, \citet{takeda2005}, \citet{luck2007}\\
68581 & 122430 & $-$0.10 & \citet{dasilva2006}\\
75458 & 137759 & $+$0.07 & \citet{santos2003}, \citet{sadakane2005}, this work\\
92895 & 175541 & $-$0.18 & \citet{valenti2005}\\
93746 & 177830 & $+$0.35 & \citet{santos2003}, \citet{valenti2005}\\
99894 & 192699 & $-$0.26 & \citet{johnson2007}\\
109577 & 210702 & $+$0.01 & \citet{luck2007}, \citet{johnson2007}\\
114855 & 219449 & 0.00 & \citet{santos2005}, \citet{sadakane2005}, \citet{luck2007}, this work\\
116727 & 222404 & $+$0.21 & \citet{luck2007}\\
NGC 2423-3 & BD-13 2130 & $+$0.06 & \citet{twarog1997}\\
\hline
\hline
\end{tabular}
\end{minipage}
\end{table*}

\begin{figure}
\includegraphics[width=\linewidth]{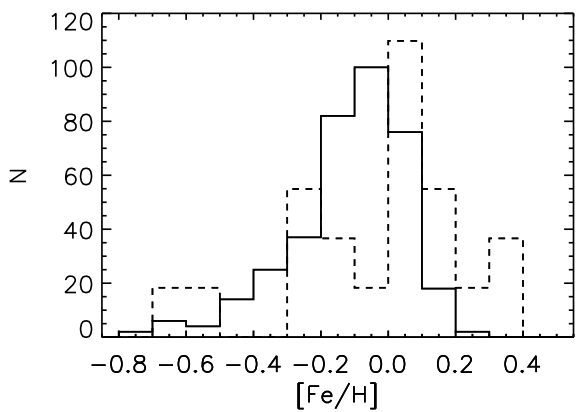}
\caption{Distribution of metallicities of all stars in our sample with mean, median and trimean values of $-$0.12, $-$0.09 and $-$0.095 dex, respectively. The metallicity of 20 giants with an announced companion in the literature (see text for selection criteria) are plotted with the dashed histogram. The latter distribution is normalised to the total number of giants in our sample. These giants have mean, median and trimean metallicity values of $-$0.05, $+$0.025 and $-$0.015 dex, respectively.} 
\label{Fehist}
\end{figure}

By now 20 sub-stellar companions are announced as orbiting giant stars. Some recent publications, e.g. \citet{sadakane2005} and \citet{pasquini2007}, suggest that giant stars with companions are metal poor, which is quite different from the known metallicity enhancement in dwarf stars hosting companions. \citet{schuler2005} and \citet{dasilva2006} argue that giant stars with companions may be metal poor, due to a stellar mass-companion relation instead of a metallicity-companion relation. Indeed, \citet{fischervalenti2005} also find a relation between stellar mass and companions, but conclude that this is likely spurious.  Here, we look at the metallicities of the giants with announced companions and compare these with the metallicities of the giants in our sample. For consistency with our work, the companion hosting giants are selected to encompass the same range in stellar parameters as our sample of giants, i.e. stars with T$_{\rm eff} \approx 4600 \pm 750$ K and $\log$g $\approx 2.1 \pm 1.5$.

The companion hosting giants and their metallicities adopted for the comparison are listed in Table~\ref{plgiants}. In order to perform a homogeneous comparison between the metallicities of the total and the companion hosting giant sample we determined the metallicity zero-points of the literature works we are using for the companion hosting giant stars. We found the following zero-points: $-0.05$ dex for \citet{luck2007}, $-0.10$ dex for \citet{schuler2005}, $-0.02$ dex for \citet{dasilva2006} and \citet{dollinger2007}, $-0.07$ dex for \citet{santos2003,santos2005} and \citet{sousa2006}, $-0.11$ dex for \citet{johnson2007}, \citet{valenti2005} and \citet{robinson2007b}, $0.00$ dex for \citet{mishenina2006}, $-0.05$ dex for \citet{sadakane2005}, \citet{sato2007} and \citet{takeda2005}, and $-0.85$ dex for \citet{twarog1997}. Note that \citet{twarog1997} only provide a mean metallicity for the cluster NGC 2423 based on photometry. The normalisation factor is based on the Hyades. For \citet{niedzielski2007} we do not have any stars in common, so we adopted a zero-point of $+0.01$ dex, which we obtained from the global comparison with the literature. In the event that more individual measurements of a star were available in the literature, the mean metallicity was adopted. The literature sources are mentioned in the last column of Table~\ref{plgiants}.

In Fig.~\ref{Fehist} the metallicity distribution of all stars in the present sample is shown together with the metallicity distribution of giant stars with announced companions. The companion hosting giant star distribution shows a gap at $-$0.1 dex. We do not know whether this gap is real or due to low number statistics.
The mean, median and trimean metallicities are $-$0.12, $-$0.09 and $-$0.095 dex and $-$0.05, $+$0.025 and $-$0.015 dex for the total and companions hosting sample, respectively, while the peaks of the histograms are at $-$0.05 dex and $+$0.05 dex. Gaussians fitted through the two distribution have their centres at $-$0.06 dex and $+$0.09 dex ($+$0.085 dex in case one Gauss is fitted and $+$0.09 for 2 Gaussian fits)  for the total and companion hosting sample, respectively. Therefore, the metallicity enhancement for companion hosting giants is $0.13 \pm 0.03$ dex. This is similar to the metallicity enhancement found by \citet{fischervalenti2005}. Their comparison between metallicities of all stars in the sample and companion hosting dwarfs reveals that companion hosting dwarfs are more metal rich by 0.13 dex. If they compare the metallicity enhancement as a function of stellar mass, they find also that, independent of mass, the metallicity distribution of dwarfs with companions is 0.12 dex higher than the average metallicity of all stars in the sample. 

Although the determination of precise stellar masses, for both our sample of field giant stars and the companion hosting giant stars, is beyond the scope of the present paper, we stress that overall both samples have comparable masses. Using the masses given by \citet{allende1999}, we estimate a typical mass of 2.0 M$_{\odot}$ for our sample, with the bulk of the field giant sample in the range 1.4 - 2.9 M$_{\odot}$ (first to third quartiles), which is the typical range covered by companion hosting giants according to \citet{sadakane2005} and \citet{johnson2007}.

The metallicity enhancement for companion hosting giants should be taken with caution. First, it is still based on low number statistics. Second, for nearly 200 stars in our sample, we do not have a long enough time span of radial velocity observations to detect companions, in case these are present. Third, there is still some discussion ongoing about some of the companion hosting giants with companions in (nearly) circular orbits. The observed radial velocity variations could in principle also be due to a mechanism intrinsic to the star.

The present conclusion that companion hosting giants are on average metal rich is rather different compared to other studies. \citet{sadakane2005} and \citet{pasquini2007} both agree that the giant stars with companions are typically not much more metal rich than [Fe/H] = 0.0. \citet{sadakane2005} analysed only a few giants, therefore their results may be due to statistics based on small numbers. However, \citet{pasquini2007} used a total of 14 giants, slightly less than our sample of 20 giants. The two samples have only 10 stars in common (HD11977, HD13189, HD28305, HD47536, HD62509, HD73108, HD104985, HD122430, HD137759, HD222404), which include the most metal poor giants in the sample, but does not contain some of the more metal rich ones. Since their histogram for the companion hosting giants does not include stars with [Fe/H] $>$ $+$0.2 dex, we suspect that the four candidates which are not available in the literature are also not as metal rich as for example HD27442 and HD177830, both with [Fe/H] $>$ $+$0.3 (\citet{santos2003,valenti2005}). Note that \citet{pasquini2007} did not correct for potential systematic errors in the literature [Fe/H] values for giant stars with companions.

Finally, caution should be taken when comparing dwarfs with giants, since systematic errors are expected due to differences in their atmospheric structures and parameters \citep{asplund2005}. Indeed, in a large study of several hundreds of giants and dwarfs, \citet{luck2007} show that the abundances of Fe, Mn and Ba in giants may be affected by systematic uncertainties when compared with dwarf stars. A consistent study such as ours, comparing field giants to companion hosting giants, is therefore more reliable.

\section{Rotational velocity}

\begin{figure}
\includegraphics[width=\linewidth]{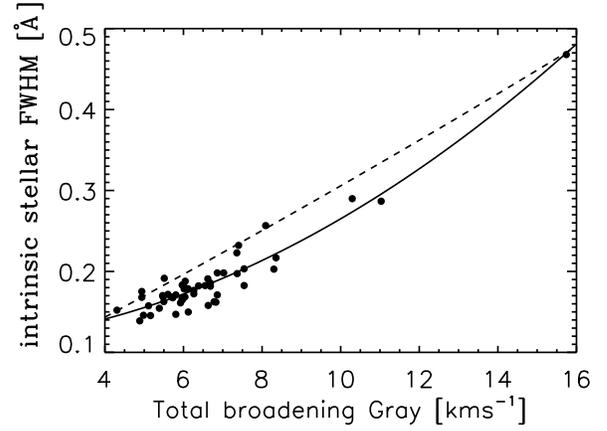}
\caption{The intrinsic stellar FWHM of spectral lines as a function Gray's total broadening, for 51 stars in common between our sample and \citet{gray1989}. The best fit (Equation~[\ref{eqFWHMcal}]) is shown as the solid line, while the best fit obtained by \citet{fekel1997} is shown as the dashed line.}
\label{FWHMcal}
\end{figure}

We computed rotational velocities for our sample of giant stars, using the method described by \citet{fekel1997}. The FWHM for moderate spectral lines at 6432.68, 6452.33, 6454.99, 6455,60, 6456.38, 6469.15 and 6471.66 \AA~is determined and averaged. The dispersion of the FWHM resulting from individual lines is typically 0.025 m\AA.
The instrumental broadening is determined from Thorium-Argon (ThAr) images taken at the beginning and end of each night.  The FWHM of several ThAr lines, in the same spectral region as the stellar lines, are determined and averaged.
The intrinsic stellar broadening is computed as $\textrm{FWHM}_{intrinsic}=\sqrt{\textrm{FWHM}_{measured}^{2}-\textrm{FWHM}_{instrumental}^{2}}$.

The intrinsic stellar broadening is converted to rotational velocity $\varv \sin i$, using the results from \citet{gray1989}. For the 51 stars in common \citep[excluding 2 outliers, for which we find higher velocities than][] {gray1989}, the intrinsic broadening is plotted as a function of the total broadening ($\sqrt{(\varv \sin i) ^{2} +\varv_{macro}^{2}}$) determined by \citet{gray1989}, as shown in Fig.~\ref{FWHMcal}. A second order polynomial is fitted:

\begin{equation}
\textrm{FWHM}_{intrinsic}=0.10963+0.002758\textrm{X}+0.001278\textrm{X}^{2},
\label{eqFWHMcal}
\end{equation}
with X the value of Gray's total broadening. The dispersion of this fit is 0.015 \AA. 
This fit is used as calibration to convert the FWHM$_{intrinsic}$ in \AA~to total broadening in kms$^{-1}$. Note that we only cover a total broadening between 4 and 16 kms$^{-1}$. All stars in our sample fall in this range. Furthermore, our fit in Fig.~\ref{FWHMcal} is different from that of \citet{fekel1997}, which is shown in Fig.~\ref{FWHMcal} with the dashed line. \citet{fekel1997} covers a much wider range in total broadening and might not be sensitive to the curvature in the particular region discussed here. In this study we used Eq.~(\ref{eqFWHMcal}) to derive the total broadening in kms$^{-1}$. From this total broadening we derive the rotational velocity as $\varv \sin i = \sqrt{\textrm{FWHM}_{total}^{2}-\varv_{macro}^{2}}$.

\subsection{Macro turbulence}
\begin{figure}
\centering
\includegraphics[width=\linewidth]{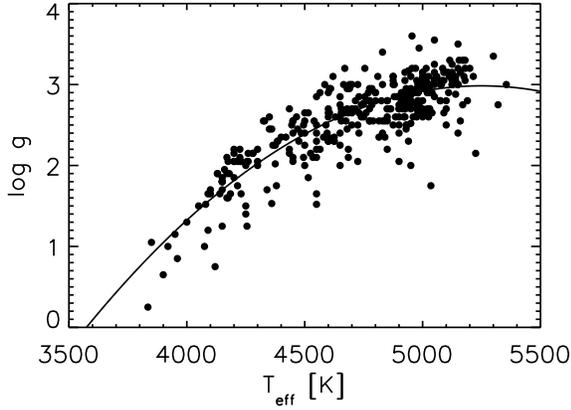}
\caption{Log~g vs. T$_{\rm{eff}}$ for all stars in our sample. The solid line shows the best fit (Eq.~[\ref{eqloggTeff}]).}
\label{loggTeff}
\end{figure}

The macro turbulence is derived from the spectral type as shown in Figure~17.10 from \citet{gray2005}. Each luminosity class has its own relation. We estimate the luminosity class from a $\log$~g vs. T$_{\rm{eff}}$ relation (Fig.~\ref{loggTeff}). Most stars in the sample are luminosity class III stars, and therefore, the second order best fit relation, shown in Eq.~(\ref{eqloggTeff}), is used for class III stars. This relation has a robust sigma scatter of 0.25 dex.
\begin{equation}
\log g_{III} = -26.332+1.117\cdot10^{-2}\textrm{T}_{\rm{eff}}-1.064\cdot10^{-6}\textrm{T}_{\rm{eff}}^{2}.
\label{eqloggTeff}
\end{equation}
Stars within a factor of 2 of the $\log g_{III}$ relation are considered to be class III giants, resulting in the following subdivision:
\begin{eqnarray}
\textrm{giants:} & & \log g = \log g_{III} \pm 0.3 \textrm{ dex}, \nonumber\\
\textrm{subgiants:} & & \log g > \log g_{III} + 0.3 \textrm{ dex}, \nonumber\\
\textrm{luminous giants:} & & \log g < \log g_{III} - 0.3 \textrm{ dex}. \nonumber
\end{eqnarray}

With the luminosity classes, we used Figure~17.10 from \citet{gray2005} to determine relations between $\varv_{macro}$ and T$_{\rm{eff}}$ for luminosity classes II, III and IV. We found the following relations:
\begin{eqnarray}
\textrm{class II:   }  \varv_{macro} = -0.214 + 0.00158 \textrm{T}_{\rm{eff}} & & \sigma = 0.55 \textrm{ kms}^{-1}\\
\textrm{class III:  }  \varv_{macro} = -3.953 + 0.00195 \textrm{T}_{\rm{eff}} & & \sigma = 0.45 \textrm{ kms}^{-1}\\
\textrm{class IV:  }  \varv_{macro} = -8.426 + 0.00241 \textrm{T}_{\rm{eff}} & & \sigma = 0.23 \textrm{ kms}^{-1}
\end{eqnarray}
In the case that $\varv_{macro}$ appeared to be higher than the total broadening, we used $\varv_{macro}$ from a higher luminosity class to determine $\varv \sin i$. In case $\varv_{macro}$ was still too high, we adopted $\varv_{macro}$ = 3 kms$^{-1}$, as used by \citet{fekel1997} for G and K giants. 

\subsection{Comparison with the literature}
\begin{figure}
\centering
\includegraphics[width=\linewidth]{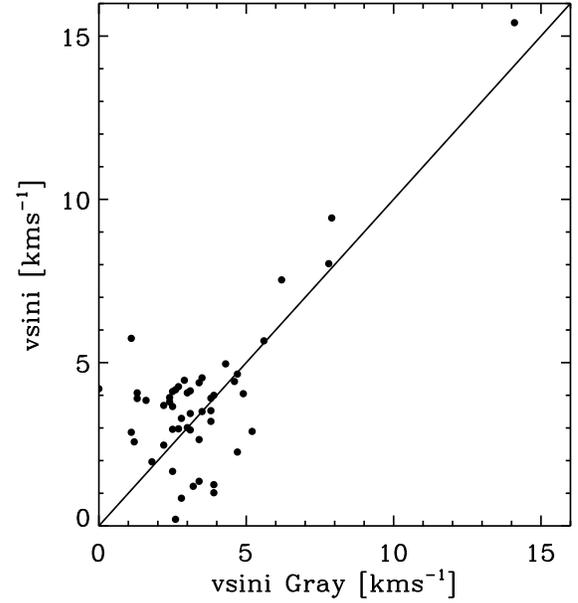}
\caption{$\varv \sin i$ obtained here vs. $\varv \sin i$ obtained by \citet{gray1989}. The solid line is a 1 to 1 relation.}
\label{vsini}
\end{figure}

\begin{figure}
\centering
\includegraphics[width=\linewidth]{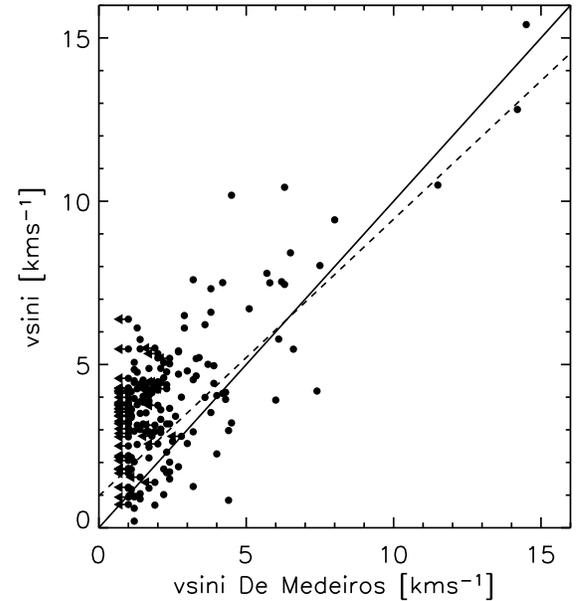}
\caption{$\varv \sin i$ obtained here vs. $\varv \sin i$ obtained by \citet{demedeiros1999}. The solid line is a 1 to 1 relation and the dashed line is the relation between $\varv \sin i$ obtained by \citet{gray1989} and \citet{demedeiros1999}. The arrows indicate upper limits.}
\label{vsiniDeM}
\end{figure}

We checked our final $\varv \sin i$ values by comparing the values of the 51 stars in common between our sample and \citet{gray1989}, see Fig.~\ref{vsini}. The values are located around the 1 to 1 relation indicated by the solid line, which shows that the results of both samples are consistent. We also have 184 stars in common with \citet{demedeiros1999} and compare our $\varv \sin i$ values with theirs in Fig.~\ref{vsiniDeM}. Our values are on average higher than those obtained by \citet{demedeiros1999}. This is probably due to the different diagnostics used. \citet{demedeiros1999} show that the relation between their $\varv \sin i$ values, and those obtained by \citet{gray1989} for class III and IV, has an offset of 1.15 and a correlation coefficient of 1.18. We plotted the 1 to 1 relation, solid line, as well as the relation between $\varv \sin i$ obtained by \citet{gray1989} and \citet{demedeiros1999} in Fig.~\ref{vsiniDeM}, dashed line. The data are located around this latter relation. This indicates that the difference between the results obtained here and from \citet{demedeiros1999} are due to the different diagnostics used to determine $\varv \sin i$. Also, \citet{luck2007} find that the CORAVEL $\varv \sin i$ values may suffer from systematic differences with respect to values derived from other techniques. For all stars $\varv \sin i$ and $\varv_{macro}$ are listed in Table~\ref{parameters} (only available in the online version).
 
\section{Summary}
We have determined spectroscopic stellar parameters for a sample of about 380 G and K giant stars. Among these, 112 stars are analysed spectroscopically for the first time. Our metallicities agree with values found in the literature and we conclude that our metallicity scale is not severely affected by systematic errors. Our temperatures are $\sim$ 50 K higher compared to those from the literature. The difference is largest for stars with lowest temperatures. This is probably due to the lower accuracy of atmosphere models in this temperature range, the increased number and strength of spectral lines and increasing dependence on micro turbulence in cooler stars. An increase in temperature causes an increase in surface gravity and our values are 0.15 dex higher compared to the literature values. 

The comparison between the mean metallicity of our total sample of giant stars and giant stars with announced companions reveals that the companion hosting stars have a 0.13 $\pm$ 0.03 dex higher metallicity than the mean metallicity of our total sample. This is in agreement with the enhanced metallicity of companion hosting dwarf stars, but is based on low number statistics.

Rotational velocities are determined using the method described by \citet{fekel1997}. Stars in common between our sample and that observed by \citet{gray1989} are used to convert FWHM of moderate lines [\AA] to total line broadening [kms$^{-1}$]. We used a $\log$~g vs. T$_{\rm{eff}}$ correlation to determine the luminosity class of the stars. This luminosity class was subsequently used to calculate the macro turbulence, which has a different relation with temperature for different classes.
Our data are in agreement with those obtained by \citet{gray1989}, but are on average larger than the values obtained by \citet{demedeiros1999}. This is due to the different diagnostics used to determine $\varv \sin i$.\\
\\

\begin{acknowledgement}
We are very grateful to Andreas Quirrenbach, Sabine Reffert and Dave Mitchell, who started the radial velocity survey of giant stars, to allow us to use the template spectra for the present analysis. We also want to thank Conny Aerts and Ignas Snellen for carefully reading the manuscript and useful discussions. We would like to thank the referee for carefully reading the manuscript and constructive comments.
\end{acknowledgement}

\bibliographystyle{aa}
\bibliography{bibpropred}

\begin{thebibliography}{55}
\expandafter\ifx\csname natexlab\endcsname\relax\def\natexlab#1{#1}\fi

\bibitem[{{Allende Prieto} \& {Lambert}(1999)}]{allende1999}
{Allende Prieto}, C. \& {Lambert}, D.~L. 1999, \aap, 352, 555

\bibitem[{{Asplund}(2005)}]{asplund2005}
{Asplund}, M. 2005, \araa, 43, 481

\bibitem[{{Bard} \& {Kock}(1994)}]{bard1994}
{Bard}, A. \& {Kock}, M. 1994, \aap, 282, 1014

\bibitem[{{Benz} \& {Mayor}(1981)}]{benz1981}
{Benz}, W. \& {Mayor}, M. 1981, \aap, 93, 235

\bibitem[{{Blackwell} {et~al.}(1995){Blackwell}, {Lynas-Gray}, \&
  {Smith}}]{blackwell1995}
{Blackwell}, D.~E., {Lynas-Gray}, A.~E., \& {Smith}, G. 1995, \aap, 296, 217

\bibitem[{{Butler} {et~al.}(1996){Butler}, {Marcy}, {Williams}, {McCarthy},
  {Dosanjh}, \& {Vogt}}]{butler1996}
{Butler}, R.~P., {Marcy}, G.~W., {Williams}, E., {et~al.} 1996, \pasp, 108, 500

\bibitem[{{Castelli} {et~al.}(1997){Castelli}, {Gratton}, \&
  {Kurucz}}]{castelli1997}
{Castelli}, F., {Gratton}, R.~G., \& {Kurucz}, R.~L. 1997, \aap, 324, 432

\bibitem[{{Cayrel de Strobel} {et~al.}(2001){Cayrel de Strobel}, {Soubiran}, \&
  {Ralite}}]{cayreldestrobel2001}
{Cayrel de Strobel}, G., {Soubiran}, C., \& {Ralite}, N. 2001, \aap, 373, 159

\bibitem[{{da Silva} {et~al.}(2006){da Silva}, {Girardi}, {Pasquini},
  {Setiawan}, {von der L{\"u}he}, {de Medeiros}, {Hatzes}, {D{\"o}llinger}, \&
  {Weiss}}]{dasilva2006}
{da Silva}, L., {Girardi}, L., {Pasquini}, L., {et~al.} 2006, \aap, 458, 609

\bibitem[{{de Medeiros} \& {Mayor}(1999)}]{demedeiros1999}
{de Medeiros}, J.~R. \& {Mayor}, M. 1999, \aaps, 139, 433

\bibitem[{{Doellinger} {et~al.}(2007){Doellinger}, {Hatzes}, {Pasquini},
  {Guenther}, {Hartmann}, {Girardi}, \& {Esposito}}]{dollinger2007}
{Doellinger}, M., {Hatzes}, A., {Pasquini}, L., {et~al.} 2007, in astro-ph,
  3672 (2007), 3672

\bibitem[{{Fekel}(1997)}]{fekel1997}
{Fekel}, F.~C. 1997, \pasp, 109, 514

\bibitem[{{Fischer} {et~al.}(2005){Fischer}, {Laughlin}, {Butler}, {Marcy},
  {Johnson}, {Henry}, {Valenti}, {Vogt}, {Ammons}, {Robinson}, {Spear},
  {Strader}, {Driscoll}, {Fuller}, {Johnson}, {Manrao}, {McCarthy},
  {Mu{\~n}oz}, {Tah}, {Wright}, {Ida}, {Sato}, {Toyota}, \&
  {Minniti}}]{fischer2005a}
{Fischer}, D.~A., {Laughlin}, G., {Butler}, P., {et~al.} 2005, \apj, 620, 481

\bibitem[{{Fischer} \& {Valenti}(2005)}]{fischervalenti2005}
{Fischer}, D.~A. \& {Valenti}, J. 2005, \apj, 622, 1102

\bibitem[{{Frink} {et~al.}(2002){Frink}, {Mitchell}, {Quirrenbach}, {Fischer},
  {Marcy}, \& {Butler}}]{frink2002}
{Frink}, S., {Mitchell}, D.~S., {Quirrenbach}, A., {et~al.} 2002, \apj, 576,
  478

\bibitem[{{Frink} {et~al.}(2001){Frink}, {Quirrenbach}, {Fischer}, {R{\"o}ser},
  \& {Schilbach}}]{frink2001}
{Frink}, S., {Quirrenbach}, A., {Fischer}, D., {R{\"o}ser}, S., \& {Schilbach},
  E. 2001, \pasp, 113, 173

\bibitem[{{Gonzalez}(1997)}]{gonzalez1997}
{Gonzalez}, G. 1997, \mnras, 285, 403

\bibitem[{{Gray}(1989)}]{gray1989}
{Gray}, D.~F. 1989, \apj, 347, 1021

\bibitem[{{Gray}(2005)}]{gray2005}
{Gray}, D.~F. 2005, {The Observation and Analysis of Stellar Photospheres} (The
  Observation and Analysis of Stellar Photospheres, 3rd Edition, by D.F.~Gray.~
  ISBN 0521851866.~Cambridge, UK: Cambridge University Press, 2005.)

\bibitem[{{Gustafsson} {et~al.}(2003){Gustafsson}, {Edvardsson}, {Eriksson},
  {Mizuno-Wiedner}, {J{\o}rgensen}, \& {Plez}}]{gustafsson2003}
{Gustafsson}, B., {Edvardsson}, B., {Eriksson}, K., {et~al.} 2003, in
  Astronomical Society of the Pacific Conference Series, Vol. 288, Stellar
  Atmosphere Modeling, ed. I.~{Hubeny}, D.~{Mihalas}, \& K.~{Werner}, 331

\bibitem[{{Hekker} {et~al.}(2006){Hekker}, {Reffert}, {Quirrenbach},
  {Mitchell}, {Fischer}, {Marcy}, \& {Butler}}]{hekker2006a}
{Hekker}, S., {Reffert}, S., {Quirrenbach}, A., {et~al.} 2006, \aap, 454, 943

\bibitem[{{Hekker} {et~al.}(2007){Hekker}, {Snellen}, {Aerts}, {Quirrenbach},
  {Reffert}, {Fischer}, {Marcy}, {Butler}, \& {Mitchell}}]{hekker2007}
{Hekker}, S., {Snellen}, I., {Aerts}, C., {et~al.} 2007, \aap, in preparation

\bibitem[{{Hinkle} {et~al.}(2000){Hinkle}, {Wallace}, {Valenti}, \&
  {Harmer}}]{hinkle2000}
{Hinkle}, K., {Wallace}, L., {Valenti}, J., \& {Harmer}, D. 2000, {Visible and
  Near Infrared Atlas of the Arcturus Spectrum 3727-9300 A} (Visible and Near
  Infrared Atlas of the Arcturus Spectrum 3727-9300 A ed.~Kenneth Hinkle, Lloyd
  Wallace, Jeff Valenti, and Dianne Harmer.~(San Francisco: ASP) ISBN:
  1-58381-037-4, 2000.)

\bibitem[{{Johnson} {et~al.}(2007){Johnson}, {Fischer}, {Marcy}, {Wright},
  {Driscoll}, {Butler}, {Hekker}, {Reffert}, \& {Vogt}}]{johnson2007}
{Johnson}, J.~A., {Fischer}, D.~A., {Marcy}, G.~W., {et~al.} 2007, \apj, 665,
  785

\bibitem[{{Luck} \& {Heiter}(2007)}]{luck2007}
{Luck}, R.~E. \& {Heiter}, U. 2007, \aj, 133, 2464

\bibitem[{{Marcy} \& {Butler}(1992)}]{marcy1992}
{Marcy}, G.~W. \& {Butler}, R.~P. 1992, \pasp, 104, 270

\bibitem[{{May} {et~al.}(1974){May}, {Richter}, \& {Wichelmann}}]{may1974}
{May}, M., {Richter}, J., \& {Wichelmann}, J. 1974, \aaps, 18, 405

\bibitem[{{Mel{\'e}ndez} \& {Barbuy}(1999)}]{melendez1999}
{Mel{\'e}ndez}, J. \& {Barbuy}, B. 1999, \apjs, 124, 527

\bibitem[{{Mel{\'e}ndez} {et~al.}(2006){Mel{\'e}ndez}, {Shchukina},
  {Vasiljeva}, \& {Ram{\'{\i}}rez}}]{melendez2006}
{Mel{\'e}ndez}, J., {Shchukina}, N.~G., {Vasiljeva}, I.~E., \&
  {Ram{\'{\i}}rez}, I. 2006, \apj, 642, 1082

\bibitem[{{Milford} {et~al.}(1994){Milford}, {O'Mara}, \& {Ross}}]{milford1994}
{Milford}, P.~N., {O'Mara}, B.~J., \& {Ross}, J.~E. 1994, \aap, 292, 276

\bibitem[{{Mishenina} {et~al.}(2006){Mishenina}, {Bienaym{\'e}}, {Gorbaneva},
  {Charbonnel}, {Soubiran}, {Korotin}, \& {Kovtyukh}}]{mishenina2006}
{Mishenina}, T.~V., {Bienaym{\'e}}, O., {Gorbaneva}, T.~I., {et~al.} 2006,
  \aap, 456, 1109

\bibitem[{{Niedzielski} {et~al.}(2007){Niedzielski}, {Konacki}, {Wolszczan},
  {Nowak}, {Maciejewski}, {Gelino}, {Shao}, {Shetrone}, \&
  {Ramsey}}]{niedzielski2007}
{Niedzielski}, A., {Konacki}, M., {Wolszczan}, A., {et~al.} 2007, ArXiv
  e-prints, 705

\bibitem[{{O'Brian} {et~al.}(1991){O'Brian}, {Wickliffe}, {Lawler}, {Whaling},
  \& {Brault}}]{obrian1991}
{O'Brian}, T.~R., {Wickliffe}, M.~E., {Lawler}, J.~E., {Whaling}, J.~W., \&
  {Brault}, W. 1991, Journal of the Optical Society of America B Optical
  Physics, 8, 1185

\bibitem[{{Pasquini} {et~al.}(2007){Pasquini}, {Doellinger}, {Weiss},
  {Girardi}, {Chavero}, {Hatzes}, {da Silva}, \& {Setiawan}}]{pasquini2007}
{Pasquini}, L., {Doellinger}, M.~P., {Weiss}, A., {et~al.} 2007, ArXiv
  e-prints, 707

\bibitem[{{Perryman} \& {ESA}(1997)}]{esa1997}
{Perryman}, M.~A.~C. \& {ESA}. 1997, {The HIPPARCOS and TYCHO catalogues.
  Astrometric and photometric star catalogues derived from the ESA HIPPARCOS
  Space Astrometry Mission} (The Hipparcos and Tycho catalogues.~Astrometric
  and photometric star catalogues derived from the ESA Hipparcos Space
  Astrometry Mission, Publisher: Noordwijk, Netherlands: ESA Publications
  Division, 1997, Series: ESA SP Series vol no: 1200, ISBN: 9290923997 (set))

\bibitem[{{Ram{\'{\i}}rez} \& {Mel{\'e}ndez}(2005)}]{ramirez2005a}
{Ram{\'{\i}}rez}, I. \& {Mel{\'e}ndez}, J. 2005, \apj, 626, 446

\bibitem[{{Reffert} {et~al.}(2006){Reffert}, {Quirrenbach}, {Mitchell},
  {Albrecht}, {Hekker}, {Fischer}, {Marcy}, \& {Butler}}]{reffert2006}
{Reffert}, S., {Quirrenbach}, A., {Mitchell}, D.~S., {et~al.} 2006, \apj, 652,
  661

\bibitem[{{Robinson} {et~al.}(2007{\natexlab{a}}){Robinson}, {Ammons},
  {Kretke}, {Strader}, {Wertheimer}, {Fischer}, \& {Laughlin}}]{robinson2007}
{Robinson}, S.~E., {Ammons}, S.~M., {Kretke}, K.~A., {et~al.}
  2007{\natexlab{a}}, \apjs, 169, 430

\bibitem[{{Robinson} {et~al.}(2007{\natexlab{b}}){Robinson}, {Laughlin},
  {Vogt}, {Fischer}, {Butler}, {Marcy}, {Henry}, {Driscoll}, {Takeda}, \&
  {Johnson}}]{robinson2007b}
{Robinson}, S.~E., {Laughlin}, G., {Vogt}, S.~S., {et~al.} 2007{\natexlab{b}},
  ArXiv e-prints, 708

\bibitem[{{Sadakane} {et~al.}(2005){Sadakane}, {Ohnishi}, {Ohkubo}, \&
  {Takeda}}]{sadakane2005}
{Sadakane}, K., {Ohnishi}, T., {Ohkubo}, M., \& {Takeda}, Y. 2005, \pasj, 57,
  127

\bibitem[{{Santos} {et~al.}(2004){Santos}, {Israelian}, \&
  {Mayor}}]{santos2004}
{Santos}, N.~C., {Israelian}, G., \& {Mayor}, M. 2004, \aap, 415, 1153

\bibitem[{{Santos} {et~al.}(2005){Santos}, {Israelian}, {Mayor}, {Bento},
  {Almeida}, {Sousa}, \& {Ecuvillon}}]{santos2005}
{Santos}, N.~C., {Israelian}, G., {Mayor}, M., {et~al.} 2005, \aap, 437, 1127

\bibitem[{{Santos} {et~al.}(2003){Santos}, {Israelian}, {Mayor}, {Rebolo}, \&
  {Udry}}]{santos2003}
{Santos}, N.~C., {Israelian}, G., {Mayor}, M., {Rebolo}, R., \& {Udry}, S.
  2003, \aap, 398, 363

\bibitem[{{Sato} {et~al.}(2007){Sato}, {Izumiura}, {Toyota}, {Kambe}, {Takeda},
  {Masuda}, {Omiya}, {Murata}, {Itoh}, {Ando}, {Yoshida}, {Ikoma}, {Kokubo}, \&
  {Ida}}]{sato2007}
{Sato}, B., {Izumiura}, H., {Toyota}, E., {et~al.} 2007, \apj, 661, 527

\bibitem[{{Schuler} {et~al.}(2006){Schuler}, {Hatzes}, {King}, {K{\"u}rster},
  \& {The}}]{schuler2006}
{Schuler}, S.~C., {Hatzes}, A.~P., {King}, J.~R., {K{\"u}rster}, M., \& {The},
  L.-S. 2006, \aj, 131, 1057

\bibitem[{{Schuler} {et~al.}(2005){Schuler}, {Kim}, {Tinker}, {King}, {Hatzes},
  \& {Guenther}}]{schuler2005}
{Schuler}, S.~C., {Kim}, J.~H., {Tinker}, Jr., M.~C., {et~al.} 2005, \apjl,
  632, L131

\bibitem[{{Sneden}(1973)}]{sneden1973}
{Sneden}, C.~A. 1973, PhD thesis, AA(The University of Texas in Austin.)

\bibitem[{{Sousa} {et~al.}(2006){Sousa}, {Santos}, {Israelian}, {Mayor}, \&
  {Monteiro}}]{sousa2006}
{Sousa}, S.~G., {Santos}, N.~C., {Israelian}, G., {Mayor}, M., \& {Monteiro},
  M.~J.~P.~F.~G. 2006, \aap, 458, 873

\bibitem[{{Sousa} {et~al.}(2007){Sousa}, {Santos}, {Israelian}, {Mayor}, \&
  {Monteiro}}]{sousa2007}
{Sousa}, S.~G., {Santos}, N.~C., {Israelian}, G., {Mayor}, M., \& {Monteiro},
  M.~J.~P.~F.~G. 2007, \aap, 469, 783

\bibitem[{{Takeda} {et~al.}(2002){Takeda}, {Ohkubo}, \&
  {Sadakane}}]{takeda2002}
{Takeda}, Y., {Ohkubo}, M., \& {Sadakane}, K. 2002, \pasj, 54, 451

\bibitem[{{Takeda} {et~al.}(2005){Takeda}, {Sato}, {Kambe}, {Izumiura},
  {Masuda}, \& {Ando}}]{takeda2005}
{Takeda}, Y., {Sato}, B., {Kambe}, E., {et~al.} 2005, \pasj, 57, 109

\bibitem[{{Taylor}(1999)}]{taylor1999}
{Taylor}, B.~J. 1999, \aaps, 134, 523

\bibitem[{{Twarog} {et~al.}(1997){Twarog}, {Ashman}, \&
  {Anthony-Twarog}}]{twarog1997}
{Twarog}, B.~A., {Ashman}, K.~M., \& {Anthony-Twarog}, B.~J. 1997, \aj, 114,
  2556

\bibitem[{{Valenti} {et~al.}(1995){Valenti}, {Butler}, \&
  {Marcy}}]{valenti1995}
{Valenti}, J.~A., {Butler}, R.~P., \& {Marcy}, G.~W. 1995, \pasp, 107, 966

\bibitem[{{Valenti} \& {Fischer}(2005)}]{valenti2005}
{Valenti}, J.~A. \& {Fischer}, D.~A. 2005, \apjs, 159, 141

\end{thebibliography}

\begin{longtable}{llrrrrrrrrrr}
\caption{\label{parameters} Stellar parameters: Star name, V magnitude, parallax (plx) in mas with its error (e$\_$plx), B-V colour, all from the Hipparcos catalogue \citep{esa1997}, effective temperature (T$_{\rm{eff}}$) in Kelvin, surface gravitity ($\log$~g) in dex, micro turbulence ($\xi$) in kms$^{-1}$, metallicity (A(Fe)), rotational velocity ($\varv \sin i$) in kms$^{-1}$ and macro turbulence ($\varv_{macro}$) in kms$^{-1}$.}\\
\hline\hline
HIP & HD & V& plx & e$\_$plx & B-V & T$_{\rm{eff}}$ & $\log$~g & $\xi$ & A(Fe) & $\varv \sin i$ & $\varv_{macro}$ \\
 & & mag & mas & mas & mag & K & dex & kms$^{-1}$ & & kms$^{-1}$ & kms$^{-1}$ \\
\hline
\endfirsthead
\caption{continued.}\\
\hline\hline
HIP & HD & V & plx & e$\_$plx & B-V & T$_{\rm{eff}}$ & $\log$~g & $\xi$ & A(Fe) & $\varv \sin i$ & $\varv_{macro}$ \\
 & & mag & mas & mas & mag & K & dex & kms$^{-1}$ & & kms$^{-1}$ & kms$^{-1}$ \\
\hline
\endhead
\hline
\endfoot
379 & 225216 & 5.68 & 10.30 & 0.58 & 1.051 & 4775 & 2.8 & 1.75 & 7.42 &  0.71 &  5.36\\
1354 & 1239 & 5.74 & 5.08 & 0.58 & 0.898 & 5150 & 2.4 & 1.81 & 7.23 &  0.84 &  7.92\\
1562 & 1522 & 3.56 & 11.26 & 0.73 & 1.214 & 4500 & 2.25 & 2.0 & 7.52 &  3.37 &  4.82\\
2006 & 2114 & 5.77 & 5.50 & 1.03 & 0.855 & 5160 & 2.55 & 1.85 & 7.36 &  1.26 &  6.11\\
2497 & 2774 & 5.59 & 8.29 & 0.68 & 1.163 & 4550 & 2.85 & 1.8 & 7.42 &  3.66 &  2.54\\
2942 & 3421 & 5.45 & 3.19 & 0.77 & 0.886 & 5225 & 2.15 & 2.57 & 7.14 &  5.51 &  8.04\\
3031 & 3546 & 4.34 & 19.34 & 0.76 & 0.871 & 4975 & 2.6 & 1.65 & 6.88 &  4.12 &  3.56\\
3179 & 3712 & 2.24 & 14.27 & 0.57 & 1.170 & 4625 & 2.30 & 2.85 & 7.29 &  6.71 &  5.07\\
3193 & 3807 & 5.90 & 5.66 & 0.94 & 1.091 & 4625 & 2.3 & 1.76 & 7.05 &  1.81 &  5.07\\
3231 & 3817 & 5.30 & 9.47 & 0.81 & 0.891 & 5025 & 2.65 & 1.5 & 7.34 &  1.21 &  5.85\\
3419 & 4128 & 2.04 & 34.04 & 0.82 & 1.019 & 4925 & 3.05 & 2.2 & 7.40 &  4.07 &  5.65\\
3607 & 4398 & 5.49 & 9.78 & 0.71 & 0.978 & 4925 & 2.7 & 1.66 & 7.30 &  2.27 &  5.65\\
3760 & 4627 & 5.92 & 4.93 & 0.82 & 1.104 & 4600 & 2.3 & 1.78 & 7.24 &  3.96 &  2.66\\
4422 & 5395 & 4.62 & 15.84 & 0.58 & 0.957 & 4860 & 2.7 & 1.64 & 7.09 &  2.16 &  3.29\\
4510 & 5575 & 5.44 & 4.63 & 0.74 & 1.076 & 4725 & 2.05 & 2.23 & 7.24 & 11.27 &  7.25\\
4587 & 5722 & 5.62 & 10.35 & 0.96 & 0.949 & 4925 & 2.7 & 1.41 & 7.31 &  4.44 &  3.44\\
4906 & 6186 & 4.27 & 17.14 & 0.81 & 0.952 & 4900 & 2.7 & 1.6 & 7.25 &  3.54 &  3.38\\
4914 & 6203 & 5.40 & 7.95 & 0.86 & 1.106 & 4650 & 2.6 & 1.54 & 7.22 &  2.72 &  5.11\\
5364 & 6805 & 3.46 & 27.73 & 0.71 & 1.161 & 4600 & 2.9 & 1.85 & 7.56 &  3.78 &  2.66\\
5571 & 7087 & 4.66 & 7.42 & 0.68 & 1.024 & 4850 & 2.55 & 1.77 & 7.34 &  3.54 &  5.50\\
5742 & 7318 & 4.67 & 8.64 & 0.81 & 1.047 & 4815 & 2.55 & 1.92 & 7.38 &  6.22 &  5.44\\
6537 & 8512 & 3.60 & 28.48 & 0.77 & 1.065 & 4750 & 2.77 & 1.8 & 7.36 &  3.61 &  3.02\\
6732 & 8763 & 5.50 & 10.63 & 0.77 & 1.106 & 4690 & 3.0 & 1.95 & 7.48 &  4.46 &  2.88\\
6999 & 9057 & 5.27 & 11.26 & 0.77 & 0.999 & 4950 & 2.8 & 1.58 & 7.55 &  4.47 &  3.50\\
7607 & 9927 & 3.59 & 18.76 & 0.74 & 1.275 & 4375 & 2.25 & 1.85 & 7.56 &  2.78 &  4.58\\
7884 & 10380 & 4.45 & 8.86 & 0.77 & 1.347 & 4300 & 2.2 & 2.25 & 7.22 &  3.00 &  4.43\\
7906 & 10348 & 5.97 & 6.23 & 0.80 & 1.015 & 4885 & 2.6 & 1.8 & 7.43 &  5.50 &  5.57\\
8198 & 10761 & 4.26 & 12.63 & 0.86 & 0.942 & 5025 & 2.9 & 1.75 & 7.49 &  4.46 &  3.68\\
9110 & 11909 & 5.09 & 4.95 & 0.95 & 0.921 & 5025 & 2.6 & 1.8 & 7.39 &  3.33 &  7.73\\
9347 & 12274 & 3.99 & 10.84 & 0.79 & 1.554 & 4200 & 2.2 & 2.35 & 7.43 &  7.52 &  1.70\\
9631 & 12641 & 5.96 & 9.89 & 0.88 & 0.851 & 4875 & 3.1 & 1.42 & 7.34 &  7.61 &  5.55\\
9884 & 12929 & 2.01 & 49.48 & 0.99 & 1.151 & 4600 & 2.7 & 1.7 & 7.36 &  3.44 &  2.66\\
10234 & 13468 & 5.94 & 9.20 & 0.83 & 0.967 & 4925 & 2.8 & 1.44 & 7.37 &  0.19 &  5.65\\
10326 & 13692 & 5.86 & 8.17 & 0.81 & 1.006 & 4970 & 3.2 & 1.53 & 7.52 &  3.64 &  3.55\\
10642 & 14129 & 5.51 & 9.58 & 0.93 & 0.962 & 5000 & 3.05 & 1.79 & 7.42 &  4.00 &  3.62\\
10729 & 13994 & 5.99 & 4.59 & 0.69 & 1.039 & 4935 & 2.5 & 2.04 & 7.29 & 10.49 &  7.58\\
11220 & 14770 & 5.19 & 8.69 & 0.67 & 0.979 & 4985 & 2.75 & 1.68 & 7.46 &  0.69 &  5.77\\
11432 & 15176 & 5.55 & 11.39 & 1.04 & 1.114 & 4650 & 2.85 & 1.65 & 7.42 &  3.60 &  2.78\\
12093 & 16161 & 4.87 & 8.77 & 1.11 & 0.880 & 5170 & 2.75 & 1.62 & 7.31 &  4.70 &  6.13\\
13288 & 17824 & 4.76 & 17.85 & 0.69 & 0.906 & 5180 & 3.3 & 1.3 & 7.56 &  3.66 &  4.06\\
13339 & 17656 & 5.86 & 8.21 & 0.79 & 0.903 & 5150 & 3.0 & 1.37 & 7.53 &  2.91 &  6.09\\
13701 & 18322 & 3.89 & 24.49 & 0.72 & 1.088 & 4700 & 3.00 & 1.58 & 7.46 &  2.92 &  2.90\\
13905 & 18449 & 4.94 & 9.31 & 0.78 & 1.235 & 4500 & 2.65 & 1.93 & 7.42 &  3.14 &  4.82\\
13965 & 18474 & 5.47 & 5.85 & 0.75 & 0.869 & 4940 & 2.3 & 1.59 & 7.18 &  2.97 &  5.68\\
14668 & 19476 & 3.79 & 29.05 & 0.66 & 0.980 & 4950 & 3.1 & 1.65 & 7.55 &  1.79 &  3.50\\
14817 & 19656 & 4.61 & 10.69 & 0.80 & 1.115 & 4600 & 2.3 & 1.89 & 7.31 &  2.57 &  5.02\\
14838 & 19787 & 4.35 & 19.44 & 1.23 & 1.033 & 4875 & 3.05 & 1.68 & 7.59 &  2.15 &  3.32\\
15549 & 20644 & 4.47 & 5.09 & 0.90 & 1.555 & 4100 & 1.65 & 3.0 & 7.05 &  5.77 &  4.04\\
15696 & 20825 & 5.55 & 3.35 & 1.16 & 1.100 & 4775 & 2.55 & 2.18 & 7.32 &  9.43 &  5.36\\
15861 & 21017 & 5.50 & 14.18 & 0.98 & 1.190 & 4620 & 3.1 & 1.75 & 7.66 &  3.37 &  2.71\\
16335 & 21552 & 4.36 & 9.23 & 0.83 & 1.367 & 4215 & 2.05 & 1.87 & 7.29 &  2.21 &  4.27\\
16358 & 21755 & 5.93 & 6.31 & 0.96 & 0.953 & 5140 & 3.05 & 1.62 & 7.46 &  4.59 &  3.96\\
16780 & 22409 & 5.56 & 8.60 & 0.77 & 0.915 & 4980 & 2.8 & 1.51 & 7.18 &  3.33 &  3.58\\
16989 & 22675 & 5.86 & 8.35 & 0.74 & 0.980 & 5000 & 3.0 & 1.34 & 7.60 &  4.41 &  3.62\\
17103 & 22796 & 5.55 & 8.14 & 0.85 & 0.931 & 4990 & 3.0 & 1.6 & 7.33 &  2.15 &  5.78\\
18212 & 24240 & 5.76 & 7.58 & 0.78 & 1.040 & 4850 & 2.7 & 1.91 & 7.48 &  4.38 &  5.50\\
19009 & 25555 & 5.46 & 3.42 & 0.90 & 0.813 & 4360 & 1.53 & 1.68 & 7.16 &  5.78 &  6.67\\
19011 & 25723 & 5.62 & 8.12 & 0.84 & 1.062 & 4775 & 2.7 & 1.67 & 7.46 &  3.66 &  3.08\\
19388 & 26162 & 5.51 & 11.21 & 0.87 & 1.077 & 4800 & 2.9 & 1.7 & 7.55 &  4.00 &  3.14\\
19483 & 26409 & 5.44 & 8.65 & 0.82 & 0.941 & 5000 & 2.8 & 1.66 & 7.45 &  2.52 &  5.80\\
19996 & 27179 & 5.95 & 5.92 & 0.76 & 1.078 & 4850 & 2.6 & 1.77 & 7.55 &  5.79 &  5.50\\
20241 & 27278 & 5.95 & 9.42 & 0.79 & 0.962 & 4950 & 2.95 & 1.47 & 7.40 &  3.75 &  3.50\\
20250 & 27382 & 4.97 & 9.53 & 0.92 & 1.150 & 4550 & 2.5 & 1.57 & 7.17 &  0.94 &  4.92\\
20252 & 27348 & 4.93 & 14.42 & 0.83 & 0.950 & 5050 & 3.07 & 1.33 & 7.60 &  4.14 &  3.74\\
20268 & 27497 & 5.76 & 7.62 & 0.93 & 0.914 & 5180 & 3.2 & 1.37 & 7.63 &  0.88 &  6.15\\
20455 & 27697 & 3.77 & 21.29 & 0.93 & 0.983 & 5000 & 3.0 & 1.5 & 7.58 &  2.96 &  5.80\\
20732 & 28100 & 4.69 & 7.17 & 0.81 & 0.979 & 4930 & 2.45 & 1.77 & 7.25 &  4.96 &  5.66\\
20885 & 28307 & 3.84 & 20.66 & 0.85 & 0.952 & 5000 & 3.0 & 1.45 & 7.57 &  4.38 &  3.62\\
20889 & 28305 & 3.53 & 21.04 & 0.82 & 1.014 & 4910 & 2.75 & 1.73 & 7.54 &  3.66 &  5.62\\
21248 & 29085 & 4.49 & 26.22 & 0.71 & 0.972 & 4875 & 3.1 & 1.35 & 7.29 &  2.15 &  3.32\\
21421 & 29139 & 0.87 & 50.09 & 0.95 & 1.538 & 4100 & 1.70 & 2.45 & 7.13 &  5.20 &  4.04\\
21743 & 29737 & 5.56 & 10.34 & 0.69 & 0.926 & 4980 & 2.8 & 1.42 & 7.20 &  1.48 &  3.58\\
22220 & 30138 & 5.99 & 7.36 & 0.85 & 0.934 & 4920 & 2.9 & 1.71 & 7.43 &  2.98 &  5.64\\
22860 & 31414 & 5.71 & 6.85 & 0.63 & 0.953 & 5150 & 3.0 & 1.93 & 7.57 &  5.25 &  6.09\\
23015 & 31398 & 2.69 & 6.37 & 0.96 & 1.490 & 3950 & 1.15 & 2.57 & 7.31 &  7.32 &  3.75\\
23123 & 31767 & 4.47 & 3.42 & 0.86 & 1.369 & 4250 & 1.4 & 2.35 & 7.26 &  4.17 &  6.50\\
23685 & 32887 & 3.19 & 14.39 & 0.68 & 1.460 & 4150 & 1.8 & 2.2 & 7.30 &  4.30 &  4.14\\
24294 & 33833 & 5.90 & 7.31 & 0.74 & 0.960 & 4980 & 3.0 & 1.54 & 7.46 &  3.24 &  3.58\\
24822 & 34559 & 4.96 & 15.83 & 0.86 & 0.937 & 5060 & 3.1 & 1.53 & 7.52 &  3.02 &  3.77\\
25247 & 35369 & 4.13 & 18.71 & 0.74 & 0.943 & 4950 & 2.8 & 1.46 & 7.32 &  2.72 &  3.50\\
27280 & 38527 & 5.78 & 10.88 & 0.86 & 0.888 & 5125 & 3.05 & 1.32 & 7.42 &  1.71 &  6.04\\
27483 & 38656 & 4.51 & 15.34 & 0.80 & 0.949 & 4980 & 2.9 & 1.46 & 7.37 &  4.22 &  3.58\\
27588 & 39118 & 5.97 & 2.89 & 0.83 & 0.953 & 4550 & 1.52 & 2.16 & 7.15 &  4.19 &  6.97\\
27629 & 39004 & 5.60 & 8.66 & 0.98 & 0.978 & 5000 & 3.05 & 1.66 & 7.50 &  3.80 &  5.80\\
28812 & 41361 & 5.67 & 2.96 & 0.87 & 1.047 & 4900 & 2.4 & 1.95 & 7.38 &  3.88 &  7.53\\
28814 & 41380 & 5.63 & 1.31 & 0.76 & 1.041 & 4900 & 2.5 & 2.9 & 7.20 & 12.81 &  7.53\\
29379 & 42398 & 5.83 & 4.33 & 0.83 & 1.110 & 4650 & 2.4 & 1.54 & 7.34 &  4.20 &  2.78\\
29575 & 43023 & 5.83 & 10.36 & 0.73 & 0.910 & 5140 & 3.1 & 1.41 & 7.53 &  3.85 &  3.96\\
30457 & 44951 & 5.21 & 7.76 & 0.74 & 1.230 & 4500 & 2.4 & 1.92 & 7.26 &  3.76 &  4.82\\
30720 & 45433 & 5.55 & 4.35 & 0.86 & 1.376 & 4200 & 1.85 & 2.0 & 7.41 &  4.70 &  4.24\\
31159 & 46241 & 5.88 & 6.30 & 0.94 & 0.997 & 4925 & 2.7 & 1.59 & 7.42 &  3.87 &  3.44\\
31592 & 47205 & 3.95 & 50.41 & 0.70 & 1.037 & 4830 & 3.4 & 1.45 & 7.70 &  1.15 &  3.00\\
31700 & 47442 & 4.42 & 7.03 & 0.62 & 1.137 & 4550 & 2.3 & 1.8 & 7.40 &  4.31 &  4.92\\
32249 & 48433 & 4.49 & 11.82 & 0.83 & 1.167 & 4550 & 2.2 & 1.76 & 7.29 &  2.17 &  4.92\\
32562 & 48781 & 5.22 & 7.69 & 0.78 & 1.131 & 4725 & 2.5 & 1.82 & 7.41 &  2.55 &  5.26\\
32814 & 49738 & 5.68 & 2.17 & 0.88 & 1.329 & 4300 & 2.0 & 2.35 & 7.44 &  5.60 &  4.43\\
33152 & 50877 & 3.89 & 1.65 & 0.62 & 1.740 & 3900 & 0.65 & 4.0 & 7.17 & 12.31 &  5.95\\
33160 & 50778 & 4.08 & 12.94 & 0.87 & 1.418 & 4050 & 1.5 & 1.8 & 7.10 &  4.27 &  3.94\\
33421 & 51000 & 5.91 & 8.47 & 0.92 & 0.878 & 5180 & 3.05 & 1.55 & 7.45 &  2.26 &  6.15\\
33449 & 50522 & 4.35 & 19.14 & 0.76 & 0.850 & 4775 & 2.8 & 0.94 & 7.45 &  2.94 &  5.36\\
33856 & 52877 & 3.49 & 2.68 & 0.59 & 1.729 & 3850 & 1.05 & 3.5 & 7.20 & 10.76 &  3.55\\
33914 & 52556 & 5.78 & 5.06 & 0.85 & 1.140 & 4700 & 2.65 & 2.3 & 7.41 &  3.84 &  5.21\\
34033 & 52960 & 5.14 & 4.39 & 0.72 & 1.391 & 4150 & 1.8 & 2.0 & 7.41 &  5.33 &  4.14\\
34267 & 53329 & 5.55 & 10.68 & 0.88 & 0.909 & 4950 & 2.7 & 1.62 & 7.03 &  4.26 &  3.50\\
34387 & 54079 & 5.74 & 5.74 & 0.86 & 1.176 & 4450 & 2.1 & 1.8 & 7.07 &  3.04 &  4.72\\
34693 & 54719 & 4.41 & 10.81 & 0.97 & 1.261 & 4500 & 2.55 & 1.96 & 7.63 &  3.03 &  4.82\\
34987 & 55751 & 5.36 & 4.35 & 0.89 & 1.193 & 4550 & 2.1 & 1.86 & 7.38 &  3.74 &  4.92\\
35476 & 56989 & 5.90 & 6.27 & 0.85 & 1.069 & 4790 & 2.55 & 1.43 & 7.50 &  7.79 &  5.39\\
35615 & 57478 & 5.59 & 5.86 & 0.74 & 0.971 & 5090 & 2.65 & 1.94 & 7.41 &  8.40 &  7.83\\
35907 & 57669 & 5.23 & 4.48 & 0.96 & 1.249 & 4500 & 2.0 & 2.35 & 7.42 &  3.21 &  6.90\\
36041 & 58367 & 4.99 & 3.30 & 0.88 & 0.991 & 4900 & 2.05 & 2.04 & 7.37 &  4.22 &  7.53\\
36388 & 59311 & 5.60 & 2.00 & 0.94 & 1.493 & 4225 & 2.2 & 2.3 & 7.30 &  5.35 &  1.76\\
36616 & 59686 & 5.45 & 10.81 & 0.75 & 1.126 & 4650 & 2.75 & 1.68 & 7.64 &  4.28 &  2.78\\
36848 & 60666 & 5.78 & 10.41 & 0.67 & 1.045 & 4750 & 2.6 & 1.38 & 7.47 &  4.03 &  3.02\\
36962 & 60522 & 4.06 & 13.57 & 0.87 & 1.540 & 4130 & 1.9 & 2.6 & 7.13 &  5.19 &  4.10\\
37204 & 60986 & 5.58 & 10.65 & 0.97 & 0.921 & 5200 & 3.2 & 1.46 & 7.65 &  3.44 &  4.11\\
37364 & 61774 & 5.92 & 4.53 & 0.71 & 1.158 & 4680 & 2.45 & 1.58 & 7.43 &  1.78 &  5.17\\
37447 & 61935 & 3.94 & 22.61 & 0.80 & 1.022 & 4825 & 2.8 & 1.6 & 7.50 &  0.70 &  5.46\\
37740 & 62345 & 3.57 & 22.73 & 0.83 & 0.932 & 5030 & 2.95 & 1.58 & 7.47 &  4.36 &  3.70\\
37826 & 62509 & 1.16 & 96.74 & 0.87 & 0.991 & 4925 & 3.15 & 1.65 & 7.56 &  1.67 &  3.44\\
38253 & 63752 & 5.60 & 2.32 & 1.03 & 1.446 & 4075 & 1.0 & 2.38 & 7.14 &  6.21 &  6.22\\
38375 & 64152 & 5.62 & 11.90 & 0.73 & 0.956 & 4930 & 2.85 & 1.7 & 7.41 &  2.60 &  3.46\\
38962 & 65345 & 5.30 & 12.33 & 0.96 & 0.933 & 5020 & 3.02 & 1.34 & 7.55 &  3.41 &  3.67\\
39079 & 65695 & 4.93 & 13.06 & 0.96 & 1.205 & 4470 & 2.45 & 1.7 & 7.34 &  1.85 &  4.76\\
39177 & 65759 & 5.60 & 4.52 & 1.01 & 1.317 & 4300 & 2.05 & 1.9 & 7.55 &  5.73 &  4.43\\
39191 & 65714 & 5.87 & 2.90 & 0.89 & 1.021 & 4920 & 2.6 & 1.76 & 7.55 &  2.50 &  5.64\\
40107 & 68312 & 5.36 & 10.32 & 0.87 & 0.892 & 5150 & 3.2 & 1.44 & 7.47 &  2.25 &  3.99\\
40305 & 68077 & 5.88 & 6.59 & 0.73 & 1.016 & 4940 & 2.8 & 1.82 & 7.49 &  4.15 &  5.68\\
40526 & 69267 & 3.53 & 11.23 & 0.97 & 1.481 & 4200 & 2.05 & 2.3 & 7.30 &  4.88 &  4.24\\
40866 & 69994 & 5.80 & 6.39 & 0.82 & 1.137 & 4650 & 2.6 & 1.57 & 7.42 &  3.15 &  2.78\\
41075 & 70272 & 4.25 & 8.39 & 0.79 & 1.550 & 4175 & 2.05 & 2.8 & 7.25 &  5.41 &  4.19\\
41704 & 71369 & 3.35 & 17.76 & 0.65 & 0.856 & 5190 & 2.8 & 1.84 & 7.33 &  3.93 &  6.17\\
41909 & 72292 & 5.33 & 10.46 & 0.89 & 1.252 & 4450 & 2.55 & 1.68 & 7.64 &  2.89 &  4.72\\
42008 & 72561 & 5.89 & 0.60 & 1.09 & 1.066 & 4840 & 2.35 & 2.32 & 7.33 &  3.91 &  7.43\\
42402 & 73471 & 4.45 & 9.25 & 0.94 & 1.216 & 4550 & 2.4 & 2.1 & 7.54 &  2.95 &  4.92\\
42911 & 74442 & 3.94 & 23.97 & 0.83 & 1.083 & 4730 & 2.65 & 1.55 & 7.56 &  3.78 &  2.97\\
43409 & 75691 & 4.02 & 15.63 & 0.58 & 1.272 & 4450 & 2.55 & 1.7 & 7.47 &  2.21 &  4.72\\
43531 & 75506 & 5.15 & 11.91 & 0.72 & 0.971 & 4830 & 2.55 & 1.6 & 7.15 &  4.05 &  3.21\\
43813 & 76294 & 3.11 & 21.64 & 0.99 & 0.978 & 4840 & 2.55 & 1.73 & 7.33 &  3.18 &  5.49\\
43834 & 76219 & 5.23 & 5.68 & 0.84 & 1.000 & 4950 & 2.9 & 1.99 & 7.37 & 10.43 &  5.70\\
43923 & 76291 & 5.72 & 14.21 & 0.78 & 1.125 & 4665 & 3.0 & 1.63 & 7.42 &  2.02 &  2.82\\
44154 & 76813 & 5.23 & 10.19 & 0.75 & 0.913 & 5020 & 2.9 & 1.47 & 7.45 &  1.87 &  5.84\\
44356 & 77353 & 5.64 & 5.32 & 0.85 & 1.163 & 4525 & 2.15 & 1.89 & 7.11 &  3.04 &  4.87\\
44406 & 77445 & 5.85 & 4.91 & 0.91 & 1.100 & 4760 & 2.65 & 1.78 & 7.44 &  4.34 &  3.05\\
44659 & 77996 & 4.99 & 2.69 & 0.93 & 1.189 & 4380 & 1.75 & 1.88 & 7.37 &  2.32 &  6.71\\
44818 & 78235 & 5.42 & 12.56 & 0.81 & 0.888 & 5170 & 3.3 & 1.38 & 7.50 &  5.39 &  4.03\\
44936 & 78668 & 5.76 & 7.09 & 0.93 & 0.937 & 5000 & 2.65 & 1.39 & 7.42 &  3.72 &  5.80\\
45412 & 79452 & 5.98 & 7.17 & 0.90 & 0.839 & 5100 & 2.7 & 1.9 & 6.86 & 10.18 &  5.99\\
46390 & 81797 & 1.99 & 18.40 & 0.78 & 1.440 & 4200 & 2.15 & 2.5 & 7.44 &  6.20 &  1.70\\
46652 & 82087 & 5.87 & 6.33 & 0.90 & 1.032 & 4850 & 2.8 & 1.7 & 7.46 &  0.24 &  5.50\\
46750 & 82308 & 4.32 & 9.69 & 0.89 & 1.541 & 4000 & 1.3 & 2.3 & 7.19 &  6.12 &  3.85\\
46880 & 82734 & 5.02 & 9.76 & 0.69 & 1.023 & 4980 & 2.9 & 1.95 & 7.60 &  6.60 &  5.76\\
46952 & 82635 & 4.54 & 18.52 & 0.88 & 0.914 & 5150 & 3.5 & 1.55 & 7.51 &  6.59 &  3.99\\
46982 & 82870 & 5.56 & 4.79 & 0.83 & 1.159 & 4600 & 2.6 & 1.85 & 7.46 &  4.09 &  2.66\\
47029 & 82741 & 4.81 & 14.23 & 0.81 & 0.992 & 4910 & 2.9 & 1.62 & 7.32 &  3.88 &  3.41\\
47189 & 83189 & 5.73 & 3.42 & 0.93 & 1.223 & 4450 & 2.05 & 2.04 & 7.55 &  8.88 &  4.72\\
47431 & 83618 & 3.90 & 11.83 & 0.80 & 1.313 & 4400 & 2.35 & 1.9 & 7.42 &  2.43 &  4.63\\
47570 & 83805 & 5.61 & 9.59 & 0.74 & 0.951 & 5020 & 2.9 & 1.65 & 7.43 &  4.45 &  3.67\\
47959 & 84561 & 5.67 & 4.65 & 0.92 & 1.489 & 4225 & 2.05 & 2.23 & 7.17 &  3.50 &  4.29\\
48356 & 85444 & 4.11 & 11.92 & 0.81 & 0.918 & 5090 & 3.05 & 1.67 & 7.47 &  1.67 &  5.97\\
48455 & 85503 & 3.88 & 24.52 & 0.87 & 1.222 & 4565 & 2.9 & 1.95 & 7.78 &  5.06 &  2.58\\
48734 & 86080 & 5.85 & 4.84 & 0.78 & 1.129 & 4650 & 2.25 & 1.79 & 7.24 &  1.44 &  5.11\\
48802 & 85945 & 5.97 & 6.99 & 0.64 & 0.895 & 5160 & 3.15 & 1.47 & 7.53 &  7.53 &  6.11\\
50027 & 88547 & 5.77 & 6.19 & 0.89 & 1.178 & 4375 & 2.02 & 2.0 & 6.96 &  2.78 &  4.58\\
50336 & 89024 & 5.84 & 10.35 & 0.90 & 1.206 & 4755 & 3.2 & 2.6 & 7.28 &  4.60 &  3.03\\
51069 & 90432 & 3.83 & 13.14 & 0.79 & 1.456 & 4225 & 2.1 & 2.15 & 7.37 &  5.87 &  4.29\\
51775 & 91612 & 5.07 & 10.23 & 0.78 & 0.921 & 5025 & 2.95 & 1.42 & 7.38 &  4.03 &  3.68\\
52689 & 93291 & 5.49 & 11.34 & 0.86 & 0.908 & 5080 & 3.05 & 1.45 & 7.43 &  3.67 &  3.82\\
52943 & 93813 & 3.11 & 23.54 & 0.81 & 1.232 & 4435 & 2.2 & 2.0 & 7.24 &  1.76 &  4.70\\
53229 & 94264 & 3.79 & 33.40 & 0.78 & 1.040 & 4725 & 3.0 & 1.58 & 7.38 &  1.81 &  2.00\\
53261 & 94247 & 5.12 & 4.82 & 0.62 & 1.355 & 4385 & 2.3 & 2.2 & 7.28 &  3.61 &  4.60\\
53316 & 94481 & 5.65 & 7.97 & 0.83 & 0.832 & 5355 & 3.0 & 1.58 & 7.52 &  4.00 &  4.48\\
53740 & 95272 & 4.08 & 18.71 & 1.03 & 1.079 & 4785 & 2.95 & 1.75 & 7.50 &  3.76 &  5.38\\
53781 & 95212 & 5.47 & 3.70 & 0.78 & 1.466 & 4150 & 1.85 & 2.1 & 7.26 &  3.77 &  4.14\\
54539 & 96833 & 3.00 & 22.21 & 0.68 & 1.144 & 4655 & 2.55 & 2.0 & 7.35 &  3.38 &  2.79\\
55086 & 97989 & 5.88 & 7.74 & 0.73 & 1.102 & 4755 & 2.85 & 2.98 & 7.14 &  3.58 &  3.03\\
55282 & 98430 & 3.56 & 16.75 & 0.82 & 1.112 & 4580 & 2.35 & 1.9 & 7.06 &  2.40 &  4.98\\
55650 & 99055 & 5.39 & 8.93 & 0.83 & 0.938 & 5020 & 2.7 & 1.71 & 7.34 &  4.52 &  3.67\\
55716 & 99196 & 5.80 & 6.97 & 0.85 & 1.376 & 4215 & 1.75 & 2.1 & 7.14 &  4.26 &  4.27\\
55797 & 99283 & 5.73 & 9.38 & 0.71 & 0.988 & 4930 & 2.85 & 1.67 & 7.31 &  4.28 &  3.46\\
55945 & 99648 & 4.95 & 5.25 & 0.84 & 1.000 & 4950 & 2.52 & 1.87 & 7.42 &  4.40 &  5.70\\
56647 & 100920 & 4.30 & 18.31 & 0.89 & 0.983 & 4910 & 2.8 & 1.61 & 7.33 &  4.27 &  3.41\\
57399 & 102224 & 3.69 & 16.64 & 0.60 & 1.181 & 4495 & 2.1 & 2.05 & 7.05 &  1.18 &  4.81\\
58181 & 103605 & 5.83 & 10.34 & 0.63 & 1.101 & 4630 & 2.6 & 1.9 & 7.32 &  3.18 &  5.08\\
58654 & 104438 & 5.59 & 9.01 & 0.77 & 1.019 & 4875 & 3.0 & 1.64 & 7.43 &  3.64 &  3.32\\
58948 & 104979 & 4.12 & 19.08 & 0.77 & 0.967 & 4950 & 2.77 & 1.68 & 7.07 &  1.55 &  5.70\\
59316 & 105707 & 3.02 & 10.75 & 0.71 & 1.326 & 4475 & 2.3 & 2.9 & 7.31 &  5.28 &  4.77\\
59501 & 106057 & 5.60 & 6.73 & 0.74 & 0.961 & 5000 & 2.95 & 1.68 & 7.41 &  3.19 &  3.62\\
59847 & 106714 & 4.93 & 13.12 & 0.88 & 0.957 & 4850 & 2.8 & 1.87 & 7.29 &  2.14 &  5.50\\
60202 & 107383 & 4.72 & 9.04 & 0.86 & 1.010 & 4880 & 3.0 & 1.67 & 7.25 &  0.60 &  5.56\\
60485 & 107950 & 4.76 & 8.30 & 0.58 & 0.877 & 5100 & 2.5 & 1.7 & 7.37 &  5.47 &  7.84\\
60646 & 108225 & 5.01 & 14.35 & 0.60 & 0.955 & 5050 & 3.0 & 1.5 & 7.57 &  4.39 &  3.74\\
60742 & 108381 & 4.35 & 19.18 & 0.83 & 1.128 & 4675 & 2.55 & 1.68 & 7.65 &  3.52 &  2.84\\
61420 & 109519 & 5.86 & 5.00 & 0.84 & 1.242 & 4495 & 2.5 & 2.65 & 7.30 &  5.42 &  4.81\\
61571 & 109742 & 5.70 & 6.29 & 0.85 & 1.436 & 4280 & 2.15 & 2.25 & 7.36 &  4.22 &  4.39\\
62103 & 110646 & 5.91 & 14.26 & 0.77 & 0.850 & 5000 & 3.07 & 1.29 & 7.01 &  1.59 &  3.62\\
63533 & 113095 & 5.97 & 8.14 & 0.78 & 0.971 & 4975 & 2.95 & 1.57 & 7.45 &  1.49 &  5.75\\
63608 & 113226 & 2.85 & 31.90 & 0.87 & 0.934 & 5115 & 3.1 & 1.71 & 7.58 &  1.69 &  6.02\\
64078 & 114038 & 5.15 & 10.66 & 0.84 & 1.138 & 4715 & 2.8 & 1.69 & 7.55 &  1.33 &  5.24\\
64540 & 115004 & 4.94 & 6.24 & 0.68 & 1.061 & 4730 & 2.4 & 1.91 & 7.39 &  7.50 &  5.27\\
64823 & 115478 & 5.33 & 10.94 & 1.00 & 1.304 & 4350 & 2.6 & 1.76 & 7.59 &  4.88 &  2.06\\
64962 & 115659 & 2.99 & 24.69 & 0.70 & 0.920 & 5110 & 2.9 & 1.55 & 7.52 &  3.35 &  6.01\\
65301 & 116292 & 5.36 & 10.20 & 0.73 & 0.987 & 4940 & 2.75 & 1.55 & 7.42 &  2.26 &  5.68\\
65323 & 116365 & 5.88 & 2.81 & 0.86 & 1.431 & 4180 & 1.9 & 2.06 & 7.16 &  4.87 &  4.20\\
66098 & 117818 & 5.21 & 12.36 & 0.78 & 0.964 & 4900 & 2.7 & 1.63 & 7.18 &  3.87 &  3.38\\
66320 & 118219 & 5.70 & 8.80 & 0.76 & 0.950 & 4915 & 2.7 & 1.55 & 7.24 &  1.37 &  5.63\\
66907 & 119458 & 5.98 & 6.71 & 0.76 & 0.857 & 5125 & 3.0 & 1.66 & 7.40 &  4.05 &  6.04\\
67210 & 120048 & 5.92 & 8.08 & 0.63 & 0.948 & 5100 & 3.15 & 1.48 & 7.55 &  2.58 &  5.99\\
67459 & 120477 & 4.05 & 13.29 & 0.81 & 1.520 & 4170 & 1.60 & 2.60 & 6.92 &  5.06 &  4.18\\
67545 & 120602 & 6.00 & 8.09 & 0.81 & 0.899 & 5140 & 2.88 & 1.59 & 7.35 &  3.82 &  6.07\\
68895 & 123123 & 3.25 & 32.17 & 0.77 & 1.091 & 4670 & 2.65 & 1.8 & 7.33 &  2.25 &  2.83\\
69427 & 124294 & 4.18 & 14.59 & 0.95 & 1.323 & 4175 & 1.6 & 1.9 & 7.02 &  4.13 &  4.19\\
69673 & 124897 & -0.05 & 88.85 & 0.74 & 1.239 & 4230 & 1.65 & 1.95 & 6.86 &  3.80 &  4.30\\
70469 & 126218 & 5.34 & 8.16 & 0.87 & 0.962 & 5125 & 3.0 & 1.72 & 7.64 &  4.59 &  3.93\\
70791 & 127243 & 5.58 & 10.59 & 0.61 & 0.864 & 5030 & 2.7 & 2.05 & 6.79 &  3.99 &  3.70\\
71053 & 127665 & 3.57 & 21.92 & 0.81 & 1.298 & 4385 & 2.3 & 2.17 & 7.30 &  3.14 &  4.60\\
71832 & 129312 & 4.86 & 5.66 & 0.82 & 0.992 & 4925 & 2.6 & 1.74 & 7.39 &  8.42 &  5.65\\
71837 & 129336 & 5.55 & 8.46 & 0.86 & 0.941 & 4980 & 2.9 & 1.67 & 7.22 &  3.41 &  3.58\\
72125 & 129972 & 4.60 & 14.48 & 0.79 & 0.972 & 4980 & 2.9 & 1.72 & 7.43 &  0.80 &  5.76\\
72210 & 129944 & 5.80 & 8.91 & 0.96 & 0.980 & 4865 & 2.7 & 1.8 & 7.19 &  4.02 &  3.30\\
72571 & 130694 & 4.42 & 10.68 & 0.83 & 1.366 & 4250 & 2.0 & 2.13 & 6.80 &  4.23 &  4.33\\
72934 & 131530 & 5.78 & 8.94 & 0.96 & 0.982 & 5015 & 3.2 & 1.49 & 7.53 &  0.85 &  5.83\\
73133 & 131918 & 5.48 & 6.06 & 0.79 & 1.491 & 4140 & 1.65 & 2.25 & 7.25 &  4.96 &  4.12\\
73166 & 132146 & 5.72 & 5.28 & 0.81 & 0.951 & 5075 & 2.8 & 1.84 & 7.46 &  2.83 &  5.94\\
73555 & 133208 & 3.49 & 14.91 & 0.57 & 0.956 & 5100 & 2.8 & 1.8 & 7.49 &  2.64 &  5.99\\
73620 & 133165 & 4.39 & 17.78 & 0.90 & 1.026 & 4700 & 2.7 & 1.9 & 7.19 &  2.24 &  2.90\\
73909 & 134190 & 5.24 & 12.53 & 0.53 & 0.958 & 4830 & 2.4 & 1.7 & 7.02 &  3.45 &  3.21\\
74239 & 134373 & 5.75 & 7.25 & 1.00 & 1.045 & 4900 & 2.85 & 2.0 & 7.45 &  4.26 &  5.60\\
74666 & 135722 & 3.46 & 27.94 & 0.61 & 0.961 & 4900 & 2.75 & 1.65 & 7.11 &  2.87 &  3.38\\
74732 & 135534 & 5.52 & 6.52 & 1.16 & 1.357 & 4365 & 2.25 & 1.92 & 7.56 &  4.24 &  4.56\\
75352 & 136956 & 5.72 & 5.41 & 0.77 & 1.039 & 5040 & 2.9 & 1.67 & 7.60 &  2.55 &  5.88\\
75458 & 137759 & 3.29 & 31.92 & 0.51 & 1.166 & 4605 & 2.95 & 1.73 & 7.60 &  3.93 &  2.67\\
75730 & 137744 & 5.64 & 3.59 & 0.93 & 1.545 & 4230 & 2.05 & 2.25 & 7.29 &  5.10 &  4.30\\
75944 & 138137 & 5.82 & 5.78 & 0.83 & 1.056 & 4935 & 2.75 & 1.98 & 7.45 &  3.36 &  5.67\\
76425 & 139195 & 5.26 & 13.89 & 0.70 & 0.925 & 5000 & 3.15 & 1.56 & 7.37 &  3.56 &  3.62\\
76810 & 140027 & 6.00 & 7.24 & 0.79 & 0.908 & 5215 & 3.1 & 1.61 & 7.48 &  4.50 &  4.14\\
77512 & 141714 & 4.59 & 19.71 & 0.73 & 0.794 & 5300 & 3.35 & 1.48 & 7.23 &  5.62 &  4.35\\
77738 & 142531 & 5.81 & 9.09 & 0.51 & 0.972 & 5000 & 3.15 & 1.52 & 7.58 &  3.07 &  3.62\\
77853 & 142198 & 4.13 & 20.02 & 0.88 & 1.003 & 4685 & 2.2 & 1.67 & 7.14 &  3.06 &  5.18\\
78132 & 142980 & 5.54 & 14.36 & 0.80 & 1.141 & 4610 & 2.95 & 1.76 & 7.46 &  4.58 &  2.68\\
78442 & 143553 & 5.82 & 13.62 & 0.79 & 1.003 & 4810 & 3.10 & 1.33 & 7.34 &  3.01 &  3.17\\
78990 & 144608 & 4.31 & 12.32 & 0.89 & 0.831 & 5320 & 2.75 & 1.85 & 7.44 &  3.50 &  6.42\\
79195 & 145206 & 5.39 & 6.60 & 0.91 & 1.446 & 4160 & 1.95 & 2.65 & 7.23 &  2.91 &  4.16\\
79540 & 145897 & 5.24 & 7.43 & 0.91 & 1.394 & 4350 & 2.45 & 1.87 & 7.48 &  3.98 &  2.06\\
79882 & 146791 & 3.23 & 30.34 & 0.79 & 0.966 & 4970 & 2.9 & 1.52 & 7.42 &  3.50 &  3.55\\
80331 & 148387 & 2.73 & 37.18 & 0.45 & 0.910 & 5110 & 3.15 & 1.55 & 7.48 &  3.69 &  3.89\\
80343 & 147700 & 4.48 & 18.32 & 0.89 & 0.996 & 4775 & 2.55 & 1.71 & 7.29 &  1.72 &  5.36\\
80693 & 148513 & 5.41 & 7.72 & 0.87 & 1.461 & 4200 & 2.05 & 2.15 & 7.47 &  3.21 &  4.24\\
80894 & 148786 & 4.29 & 15.53 & 0.77 & 0.924 & 5175 & 3.1 & 1.68 & 7.65 &  5.01 &  6.14\\
81660 & 151101 & 4.84 & 4.79 & 0.45 & 1.212 & 4535 & 2.1 & 2.28 & 7.36 &  1.17 &  6.95\\
81724 & 150416 & 4.91 & 8.34 & 0.85 & 1.095 & 5000 & 2.6 & 1.86 & 7.36 &  5.19 &  7.69\\
81833 & 150997 & 3.48 & 29.11 & 0.52 & 0.916 & 5020 & 3.0 & 1.43 & 7.29 &  2.47 &  5.84\\
83000 & 153210 & 3.19 & 37.99 & 0.75 & 1.160 & 4655 & 2.7 & 1.82 & 7.56 &  2.16 &  5.12\\
83254 & 153834 & 5.69 & 1.07 & 0.74 & 1.332 & 4340 & 1.7 & 2.61 & 7.34 &  5.23 &  6.64\\
84380 & 156283 & 3.16 & 8.89 & 0.52 & 1.437 & 4170 & 1.9 & 2.26 & 7.50 &  6.12 &  4.18\\
84671 & 156681 & 5.03 & 4.72 & 0.80 & 1.539 & 4170 & 2.1 & 2.25 & 7.28 &  5.47 &  1.62\\
84950 & 157681 & 5.69 & 5.52 & 0.51 & 1.463 & 4255 & 2.05 & 2.18 & 7.30 &  3.78 &  4.34\\
85139 & 157617 & 5.77 & 3.03 & 0.79 & 1.251 & 4565 & 2.2 & 2.47 & 7.46 &  7.34 &  4.95\\
85355 & 157999 & 4.34 & 2.78 & 0.92 & 1.480 & 4080 & 1.52 & 2.54 & 7.42 &  7.51 &  4.00\\
85693 & 158899 & 4.41 & 8.88 & 0.64 & 1.434 & 4325 & 2.55 & 2.62 & 7.40 &  6.22 &  2.00\\
85715 & 158974 & 5.63 & 8.65 & 0.56 & 0.960 & 5090 & 3.15 & 1.57 & 7.58 &  2.01 &  5.97\\
85888 & 159501 & 5.72 & 8.62 & 0.53 & 1.089 & 4685 & 2.65 & 1.77 & 7.20 &  0.94 &  2.86\\
86742 & 161096 & 2.76 & 39.78 & 0.75 & 1.168 & 4680 & 2.95 & 2.02 & 7.62 &  3.84 &  2.85\\
87808 & 163770 & 3.86 & 4.87 & 0.54 & 1.350 & 4255 & 1.25 & 2.75 & 7.38 &  7.45 &  6.51\\
87847 & 163532 & 5.44 & 7.66 & 0.71 & 1.162 & 4800 & 2.8 & 1.98 & 7.42 &  1.46 &  5.41\\
87933 & 163993 & 3.70 & 24.12 & 0.52 & 0.935 & 5085 & 3.20 & 1.8 & 7.49 &  4.53 &  3.83\\
88048 & 163917 & 3.32 & 21.35 & 0.79 & 0.987 & 4900 & 2.85 & 2.05 & 7.55 &  3.04 &  5.60\\
88636 & 165683 & 5.72 & 4.69 & 0.62 & 1.179 & 4600 & 2.35 & 2.05 & 7.37 &  1.39 &  5.02\\
88684 & 165438 & 5.74 & 28.61 & 0.83 & 0.968 & 4955 & 3.60 & 1.22 & 7.60 &  & \\
88765 & 165760 & 4.64 & 13.71 & 0.82 & 0.951 & 5025 & 3.0 & 1.75 & 7.45 &  1.02 &  5.85\\
88839 & 165634 & 4.55 & 9.38 & 0.77 & 0.938 & 4980 & 2.65 & 1.73 & 7.44 &  1.49 &  3.58\\
89008 & 166640 & 5.57 & 7.52 & 0.57 & 0.915 & 5080 & 3.0 & 1.54 & 7.50 &  1.80 &  5.95\\
89587 & 167768 & 5.99 & 9.91 & 0.83 & 0.890 & 4930 & 2.5 & 2.05 & 6.72 &  4.42 &  7.58\\
89826 & 168775 & 4.33 & 13.71 & 0.56 & 1.162 & 4590 & 2.50 & 1.7 & 7.64 &  2.81 &  5.00\\
89919 & B+62 1612 & 8.88 & 3.81 & 0.80 & 1.058 & 5130 & 3.1 & 1.6 & 7.41 &  & \\
89962 & 168723 & 3.23 & 52.81 & 0.75 & 0.941 & 4955 & 3.2 & 1.33 & 7.34 &  0.44 &  3.52\\
90067 & 169191 & 5.25 & 7.48 & 0.66 & 1.250 & 4515 & 2.65 & 1.75 & 7.44 &  0.95 &  4.85\\
90139 & 169414 & 3.85 & 25.40 & 0.65 & 1.168 & 4585 & 3.0 & 1.45 & 7.56 &  3.48 &  2.62\\
90496 & 169916 & 2.82 & 42.20 & 0.90 & 1.025 & 4770 & 2.9 & 1.45 & 7.44 &  3.81 &  3.07\\
91004 & 171115 & 5.49 & 0.95 & 0.97 & 1.795 & 3835 & 0.25 & 3.1 & 7.03 &  7.31 &  5.85\\
91105 & 171391 & 5.12 & 11.25 & 0.78 & 0.926 & 5125 & 3.15 & 1.55 & 7.47 &  3.05 &  6.04\\
91117 & 171443 & 3.85 & 18.72 & 0.81 & 1.317 & 4280 & 2.15 & 1.88 & 7.51 &  4.58 &  4.39\\
92747 & 174947 & 5.68 & 1.88 & 0.85 & 1.206 & 4685 & 2.10 & 2.48 & 7.38 &  3.82 &  7.19\\
93026 & 175751 & 4.83 & 15.77 & 0.89 & 1.057 & 4730 & 2.85 & 1.85 & 7.46 &  3.00 &  2.97\\
93085 & 175775 & 3.52 & 8.76 & 0.99 & 1.151 & 4595 & 2.4 & 2.1 & 7.48 &  6.29 &  5.01\\
93429 & 176678 & 4.02 & 21.95 & 0.92 & 1.079 & 4690 & 2.95 & 1.53 & 7.51 &  3.91 &  2.88\\
93864 & 177716 & 3.32 & 27.09 & 1.48 & 1.169 & 4690 & 3.2 & 3.9 & 7.22 &  1.04 &  4.90\\
94302 & 180006 & 5.13 & 9.57 & 0.47 & 1.008 & 4940 & 2.9 & 1.54 & 7.58 &  5.00 &  5.68\\
94624 & 180262 & 5.58 & 5.32 & 0.90 & 1.067 & 4960 & 2.6 & 1.77 & 7.49 &  5.48 &  5.72\\
94779 & 181276 & 3.80 & 26.48 & 0.49 & 0.950 & 5050 & 3.25 & 1.65 & 7.58 &  4.28 &  3.74\\
94820 & 180540 & 4.88 & 6.09 & 0.86 & 1.013 & 4850 & 2.2 & 1.95 & 7.31 &  4.08 &  7.45\\
95352 & 182694 & 5.85 & 8.06 & 0.47 & 0.924 & 5115 & 3.1 & 1.59 & 7.48 &  3.20 &  6.02\\
95785 & 183491 & 5.82 & 6.74 & 0.72 & 1.023 & 4890 & 2.85 & 1.64 & 7.57 &  2.80 &  5.58\\
96229 & 184406 & 4.45 & 29.50 & 0.78 & 1.176 & 4670 & 3.2 & 1.82 & 7.53 &  2.78 &  2.83\\
96327 & 184492 & 5.12 & 7.34 & 0.76 & 1.122 & 4875 & 2.5 & 2.16 & 7.33 &  8.88 &  7.49\\
96459 & 185351 & 5.17 & 24.64 & 0.49 & 0.928 & 5050 & 3.55 & 1.46 & 7.49 &  2.06 &  3.00\\
96516 & 185194 & 5.67 & 6.89 & 0.72 & 1.007 & 4975 & 2.7 & 1.74 & 7.51 &  3.32 &  5.75\\
97118 & 186675 & 4.89 & 11.70 & 0.50 & 0.948 & 5050 & 2.85 & 1.65 & 7.47 &  2.94 &  5.89\\
97402 & 187193 & 6.00 & 8.16 & 0.72 & 0.993 & 4930 & 2.95 & 1.52 & 7.38 &  1.24 &  5.66\\
98337 & 189319 & 3.51 & 11.90 & 0.71 & 1.571 & 4150 & 1.70 & 2.85 & 7.09 &  5.81 &  4.14\\
98571 & 190147 & 5.06 & 7.60 & 0.47 & 1.122 & 4700 & 2.5 & 1.95 & 7.38 &  3.63 &  5.21\\
98823 & 190327 & 5.51 & 6.19 & 0.79 & 1.063 & 4850 & 2.7 & 1.98 & 7.34 & 10.30 &  5.50\\
99951 & 192944 & 5.30 & 6.91 & 0.64 & 0.951 & 5000 & 2.7 & 1.68 & 7.39 &  5.02 &  5.80\\
100064 & 192947 & 3.58 & 30.01 & 0.91 & 0.883 & 5035 & 1.75 & 2.62 & 7.34 &  7.79 &  7.74\\
100587 & 194317 & 4.43 & 12.77 & 0.62 & 1.331 & 4435 & 2.7 & 2.0 & 7.53 &  4.85 &  2.26\\
100754 & 194577 & 5.68 & 6.00 & 0.73 & 0.921 & 5075 & 3.0 & 1.47 & 7.51 &  4.65 &  5.94\\
101870 & 196753 & 5.91 & 1.65 & 0.74 & 0.953 & 4550 & 1.65 & 2.3 & 7.13 &  6.20 &  6.97\\
101986 & 197139 & 5.97 & 6.93 & 0.54 & 1.186 & 4485 & 2.4 & 1.44 & 7.41 &  6.39 &  4.79\\
102422 & 198149 & 3.41 & 69.73 & 0.49 & 0.912 & 4985 & 3.45 & 1.34 & 7.31 &  1.04 &  2.00\\
102453 & 197912 & 4.22 & 15.84 & 0.62 & 1.051 & 4940 & 3.17 & 1.87 & 7.46 &  4.31 &  3.48\\
102488 & 197989 & 2.48 & 45.26 & 0.53 & 1.021 & 4785 & 2.75 & 1.62 & 7.38 &  3.01 &  3.11\\
102978 & 198542 & 4.12 & 5.19 & 0.95 & 1.633 & 3960 & 0.85 & 2.9 & 6.88 &  4.68 &  6.04\\
103294 & 199253 & 5.19 & 6.88 & 0.74 & 1.119 & 4625 & 2.35 & 1.75 & 7.30 &  4.80 &  5.07\\
103360 & 199612 & 5.92 & 2.56 & 0.52 & 1.054 & 4740 & 2.6 & 2.0 & 7.40 &  4.76 &  5.29\\
104060 & 200905 & 3.72 & 2.77 & 0.52 & 1.609 & 3920 & 1.00 & 3.1 & 7.08 &  9.30 &  3.69\\
104459 & 201381 & 4.50 & 19.93 & 0.77 & 0.926 & 5025 & 3.10 & 1.5 & 7.48 &  3.29 &  3.68\\
104963 & 202320 & 5.17 & 4.72 & 0.82 & 1.161 & 4515 & 1.85 & 2.07 & 7.23 &  4.14 &  6.92\\
105412 & 203222 & 5.87 & 9.69 & 0.86 & 0.912 & 5050 & 3.05 & 1.46 & 7.50 &  1.79 &  5.89\\
105497 & 203644 & 5.68 & 9.93 & 0.55 & 1.100 & 4740 & 2.75 & 1.77 & 7.53 &  4.25 &  5.29\\
105515 & 203387 & 4.28 & 15.13 & 0.80 & 0.888 & 5025 & 3.0 & 1.48 & 7.34 &  5.67 &  5.85\\
106039 & 204381 & 4.50 & 18.18 & 0.89 & 0.889 & 5155 & 3.30 & 1.78 & 7.47 &  5.74 &  4.00\\
106481 & 205435 & 3.98 & 26.20 & 0.51 & 0.885 & 5125 & 3.25 & 1.46 & 7.34 &  4.26 &  3.93\\
107188 & 206453 & 4.72 & 11.22 & 0.79 & 0.868 & 5040 & 2.65 & 1.72 & 7.10 &  3.72 &  5.88\\
107315 & 206778 & 2.38 & 4.85 & 0.84 & 1.520 & 4150 & 1.25 & 3.5 & 7.31 &  8.32 &  6.34\\
107382 & 206834 & 5.10 & 4.08 & 0.89 & 1.108 & 4815 & 2.35 & 2.63 & 7.30 &  3.75 &  7.39\\
108691 & 209128 & 5.60 & 4.93 & 0.83 & 1.279 & 4465 & 2.60 & 2.18 & 7.40 &  4.04 &  4.75\\
109023 & 209761 & 5.75 & 8.22 & 0.73 & 1.249 & 4420 & 2.35 & 1.73 & 7.41 &  5.19 &  4.67\\
109068 & 209747 & 4.86 & 12.38 & 0.90 & 1.443 & 4130 & 1.90 & 2.03 & 7.51 &  4.77 &  4.10\\
109492 & 210745 & 3.39 & 4.49 & 0.50 & 1.558 & 4120 & 0.75 & 3.4 & 7.27 & 10.64 &  6.30\\
109602 & 210762 & 5.97 & 1.07 & 0.70 & 1.500 & 4185 & 1.65 & 2.45 & 7.49 &  7.87 &  4.21\\
109754 & 211073 & 4.50 & 5.79 & 0.64 & 1.385 & 4360 & 2.45 & 2.77 & 7.40 &  6.50 &  2.08\\
109937 & 211388 & 4.14 & 5.20 & 0.61 & 1.447 & 4260 & 2.15 & 2.7 & 7.50 &  7.60 &  4.35\\
109972 & 211554 & 5.88 & 4.48 & 0.56 & 0.950 & 5075 & 2.8 & 1.84 & 7.57 &  5.21 &  5.94\\
110000 & 211361 & 5.34 & 6.74 & 0.88 & 1.132 & 4800 & 2.9 & 1.83 & 7.48 &  3.43 &  5.41\\
110003 & 211391 & 4.17 & 17.04 & 0.74 & 0.979 & 5000 & 3.1 & 1.67 & 7.56 &  3.94 &  3.62\\
110023 & 211434 & 5.75 & 9.56 & 0.86 & 0.878 & 5025 & 2.7 & 1.6 & 7.18 &  1.96 &  5.85\\
110532 & 212320 & 5.92 & 7.10 & 0.93 & 0.998 & 5030 & 2.9 & 1.95 & 7.22 &  2.89 &  5.86\\
110602 & 212430 & 5.76 & 6.01 & 0.76 & 0.970 & 4975 & 2.75 & 1.73 & 7.28 &  4.07 &  3.56\\
110986 & 213119 & 5.60 & 5.63 & 0.85 & 1.578 & 4090 & 1.65 & 2.5 & 7.01 &  5.01 &  4.02\\
111362 & 213930 & 5.72 & 9.60 & 0.53 & 0.966 & 4975 & 3.05 & 1.61 & 7.54 &  5.18 &  5.75\\
111394 & 213789 & 5.88 & 7.35 & 0.86 & 0.977 & 5015 & 3.0 & 1.66 & 7.39 &  3.80 &  5.83\\
111925 & 214878 & 5.94 & 9.49 & 0.54 & 0.946 & 5050 & 3.15 & 1.49 & 7.53 &  3.86 &  3.74\\
111944 & 214868 & 4.50 & 10.81 & 0.56 & 1.318 & 4445 & 2.50 & 2.05 & 7.32 &  1.80 &  4.71\\
112067 & 214995 & 5.92 & 12.22 & 0.79 & 1.114 & 4680 & 2.70 & 1.8 & 7.45 &  7.21 &  5.17\\
112242 & 215373 & 5.11 & 11.89 & 0.60 & 0.960 & 4950 & 2.87 & 1.69 & 7.50 &  3.61 &  5.70\\
112440 & 215665 & 3.97 & 8.26 & 0.70 & 1.070 & 4650 & 2.0 & 1.89 & 7.23 &  8.03 &  7.13\\
112529 & 215721 & 5.24 & 12.26 & 0.87 & 0.941 & 4900 & 2.6 & 1.86 & 7.01 &  3.90 &  3.38\\
112724 & 216228 & 3.50 & 28.27 & 0.52 & 1.053 & 4830 & 3.00 & 1.59 & 7.54 &  3.27 &  3.21\\
112748 & 216131 & 3.51 & 27.95 & 0.77 & 0.933 & 4980 & 2.9 & 1.51 & 7.39 &  0.20 &  5.76\\
113084 & 216646 & 5.82 & 9.63 & 0.79 & 1.136 & 4600 & 2.65 & 1.65 & 7.56 &  3.05 &  2.66\\
113562 & 217303 & 5.66 & 4.62 & 0.71 & 1.253 & 4250 & 1.50 & 1.76 & 6.75 &  3.86 &  4.33\\
113622 & 217459 & 5.85 & 5.96 & 0.80 & 1.343 & 4260 & 2.05 & 1.84 & 7.36 &  2.51 &  4.35\\
113686 & 217563 & 5.94 & 1.23 & 0.91 & 0.992 & 4950 & 2.0 & 2.27 & 7.36 &  2.14 &  7.61\\
113864 & 218029 & 5.25 & 8.48 & 0.52 & 1.248 & 4450 & 2.4 & 1.78 & 7.68 &  3.18 &  4.72\\
114341 & 218594 & 3.68 & 13.96 & 0.94 & 1.202 & 4435 & 2.15 & 2.09 & 7.29 &  3.59 &  4.70\\
114449 & 218792 & 5.68 & 6.45 & 0.81 & 1.330 & 4330 & 2.55 & 1.91 & 7.51 &  4.12 &  2.01\\
114855 & 219449 & 4.24 & 21.97 & 0.89 & 1.107 & 4715 & 2.70 & 1.73 & 7.46 &  2.78 &  5.24\\
114971 & 219615 & 3.70 & 24.92 & 0.89 & 0.916 & 4940 & 2.9 & 1.76 & 6.95 &  4.20 &  3.48\\
115152 & 219945 & 5.44 & 9.95 & 0.63 & 1.014 & 4880 & 2.85 & 1.55 & 7.40 &  1.21 &  5.56\\
115438 & 220321 & 3.96 & 20.14 & 0.72 & 1.082 & 4655 & 2.65 & 1.59 & 7.23 &  3.08 &  2.79\\
115669 & 220704 & 4.38 & 10.57 & 0.72 & 1.460 & 4150 & 1.8 & 2.34 & 7.22 &  3.90 &  4.14\\
115830 & 220954 & 4.27 & 20.54 & 0.80 & 1.062 & 4775 & 2.95 & 1.84 & 7.51 &  1.79 &  3.08\\
117375 & 223252 & 5.49 & 11.19 & 0.85 & 0.941 & 5000 & 3.0 & 1.6 & 7.41 &  2.57 &  5.80\\
117567 & 223559 & 5.70 & 7.09 & 0.92 & 1.488 & 4090 & 1.20 & 2.25 & 6.93 &  3.44 &  4.02\\
117756 & 223807 & 5.76 & 5.33 & 0.81 & 1.171 & 4605 & 2.65 & 1.89 & 7.47 &  3.60 &  2.67\\
118209 & 224533 & 4.88 & 14.58 & 0.83 & 0.930 & 5115 & 3.3 & 1.56 & 7.47 &  4.64 &  3.90\\
\end{longtable}

\end{document}